\documentclass[preprint]{aa}  

\usepackage{graphicx}
\usepackage{txfonts}

\usepackage{txfonts}

\usepackage{graphicx}	
\usepackage{amsmath}	
\usepackage{amssymb}	
\usepackage{xcolor}
\usepackage{natbib}

\newcommand{\pri}{$^\prime$}
\newcommand{\pripri}{$^{\prime\prime}$}
\newcommand{\nh}{N${\rm _H}$}

\newcommand{\msun}{$M_{\odot}$}

\newcommand{\fluxcgs}{ergs~s$^{-1}$~cm$^{-2}$}
\newcommand{\lumcgs}{ergs~s$^{-1}$}

\newcommand{\nicer}{\textit{NICER}}
\newcommand{\erosita}{\textit{eROSITA}}
\newcommand{\rosat}{\textit{ROSAT}}
\newcommand{\srgerosita}{\textit{SRG/eROSITA}}
\newcommand{\spektr}{\textit{SRG}}
\newcommand{\artxc}{\textit{ART-XC}}
\newcommand{\srgcv}{SRGA\,J213151.5+491400}
\newcommand{\gaia}{\textit{Gaia}}
\newcommand{\gcv}{\textit{Gaia19fld}}
\newcommand{\xmm}{\textit{XMM-Newton}}
\newcommand{\swi}{\textit{Swift}}
\newcommand{\integral}{\textit{INTEGRAL}}

\begin{document} 

\title{Optical identification and follow-up observations of SRGA J213151.5+491400 - a new magnetic cataclysmic variable discovered with SRG Observatory}
\titlerunning{Optical identification and follow-ups of a new MCV detected by SRG}

\author{\c{S}. Balman\inst{1,2}\thanks{E-mail: solen.balman@istanbul.edu.tr, solen.balman@gmail.com}
\and
I. Khamitov \inst{3,4}
\and
A. Kolbin \inst{5,3}
\and
E. Aktekin \c{C}al{\i}\c{s}kan \inst{6}
\and
I. Bikmaev \inst{3}
\and
A. \"Ozd\"onmez\inst{7}
\and
R. Burenin \inst{8,9}
\and
Y. K{\i}l{\i}\c{c} \inst{4,10}
\and
H. H. Eseno\u{g}lu \inst{1}
\and
K. F. Yelkenci \inst{1}
\and
D. Zengin \c{C}amurdan \inst{11}
\and
M. Gilfanov \inst{12,8}
\and
I. Nas{\i}ro\u{g}lu \inst{7}
\and
E. Sonba\c{s} \inst{13,14}
\and
M. Gabdeev \inst{5}
\and
E. Irtuganov \inst{3}
\and
A. T. Sayga\c{c} \inst{1}
\and
E. Nikolaeva \inst{3}
\and
N. Sakhibullin \inst{3}
\and
H. Er \inst{7}
\and
S. Sazonov \inst{8,9}
\and
P. Medvedev \inst{8}
\and
T. G\"uver \inst{1}
\and
S. Fi\c{s}ek \inst{1}}

\authorrunning{\c{S}\"olen Balman, et al.}

\institute{Department of Astronomy and Space Sciences, Istanbul University, Faculty of Science, Beyazit, Istanbul, 34119, Turkey
\and
Faculty of Engineering and Natural Sciences, Kadir Has University, Cibali, Istanbul, 34083,Turkey
\and
Department of Astronomy and Satellite Geodesy, Kazan Federal University, Kremlevskaya Str., 18, Kazan, 420008, Russia
\and
T\"UB{İ}TAK National Observatory, Akdeniz University Campus, Antalya, 07058, Turkey
\and
Special Astrophysical Observatory of Russian Academy of Sciences, Nizhnij Arkhyz, Karachai-Cherkessian Rep., 369167 Russia 
\and
Department of Physics, S\"uleyman Demirel University, Isparta, 32000, Turkey
\and
Department of Astronomy and Space Sciences, Ataturk University, Faculty of Science, Yakutiye, Erzurum, 25240, Turkey
\and
Space Research Institute, Russian Academy of Sciences, Profsoyuznaya ul. 84/32, Moscow, 117997 Russia
\and
National Research University Higher School of Economics, Moscow 101000, Russia
\and
Department of Space Sciences and Technologies, Akdeniz University, Faculty of Sciences, Antalya, 07058, Turkey
\and
Department of Astronomy and Space Sciences, Ege University, Science Faculty, Bornova, Izmir, 35100, Turkey
\and
Max Planck Institut f\"ur Astrophysik, Karl-Schwarzschild-Str. 1, Postfach 1317, D-85741 Garching, Germany
\and
Department of Physics, Ad{\i}yaman University, Ad{\i}yaman, 02040, Turkey
\and
Astrophysics Application and Research Center, Ad{\i}yaman University, Adiyaman, 02040, Turkey
}

   \date{Received October XX, 2023; accepted XXX XX, 2023}




 \abstract
   {The paper is comprised of optical identification and multiwavelength study of a new X-ray source discovered by the Spectrum Roentgen-Gamma (\spektr) observatory during the \artxc\ survey and its follow-up optical and X-ray observations.}
    {We identify SRGA\,J213151.5+491400 in the optical wavelengths. We determine spectra and light curves in the optical high and low states to find periodicities in the light curves and resolve emission lines in the system using optical ground-based data. We study the spectral and temporal X-ray characteristics of the new source using the \spektr\ surveys in the high and low states and \nicer\ data in the low state.}
   {We present optical data from telescopes in Turkey (RTT-150 and T100 at the T\"UB\.{I}TAK National Observatory), and in Russia (6-m and 1-m at SAO RAS), together with the X-ray data obtained with \artxc\ and \erosita\ telescopes aboard \spektr\ and the \nicer\ observatory. Using the optical data we perform astrometry, photometry, spectroscopy and power spectral analysis of the optical time series. We perform optical Doppler Tomography. We present X-ray data analysis producing light curves and spectra.}
   {We detect \srgcv\ in a high state in 2020 (17.9 mag) that decreases about 3 mag into a low state (21 mag) in 2021. We find only one significant period using optical photometric time series analysis which reveals the white dwarf spin/orbital period to be 0.059710(1) days (85.982 min). The long slit spectroscopy in the high state yields a power law continuum increasing towards the blue with a prominent He\,{\sc II} line along with the Balmer line emissions with no cyclotron humps; consistent with magnetic cataclysmic variable (MCV) nature.
   Doppler Tomography confirms the polar nature revealing ballistic stream accretion along with magnetic stream during the high state. These  characteristics show that the new source is a polar-type MCV. {\ bf \artxc\ detections  yield an X-ray flux of (4.0-7.0)$\times$10$^{-12}$ \fluxcgs\ in the high state}. \erosita\ detects a dominating hot plasma component (kT$_{\rm{max}}$ $>$ 21 keV in the high state) declining to (4.0-6.0)$\times$10$^{-13}$ \fluxcgs\ in 2021 (low state). The \nicer\ data obtained in the  low state reveal a two-pole accretor showing a soft X-ray component  at (6-7)$\sigma$ significance with a blackbody temperature of 15-18 eV. A soft X-ray component has never been detected for a polar in the low state before.}
  {}


\maketitle 



\nolinenumbers

\section{Introduction}\label{intro}

Cataclysmic Variables (CVs) and related systems (e.g., AM CVns, Symbiotics) are compact binary systems with white dwarf (WD) primaries.
CVs mainly accrete through a disk and the  donor star is a late-type main sequence star or sometimes a slightly evolved star. Systems show orbital periods typically in the 1.4-15 hrs range with few exceptions out to 2-2.5 day binaries \citep{Balman2020}. 

CVs have two main sub-classes \citep{1995Warner}. An accretion
disk forms and reaches all the way to the WD in cases where the magnetic field of the WD is weak or nonexistent ($\rm{B}$ $<$ 0.01 MG), such systems are referred as nonmagnetic CVs and are characterized by their eruptive behavior \citep[][for a recent review]{1995Warner,Balman2020}.
The other class is the MCVs, divided into two sub-classes according to the
degree of synchronization of the binary widely studied in the X-rays comprising about 25\% of the CV population. Polars have strong magnetic fields in the range about 10-230 MG \citep{2020Ferrario,deMartinoetal2020},
which cause the accretion flow to directly channel onto the magnetic pole/s of the WD inhibiting the formation of an accretion disk \citep[see][for a review]{Mukai2017}.
The magnetic and tidal torques cause the WD rotation to synchronize with the binary orbit. 
Among the subclass of Polars, there are about 8 slightly asynchronous systems with $|P_{\rm spin}-P_{\rm orb}|$/P$_{\rm orb}$$\sim$ 1-3\% \citep{Schwarzetal2007,Tovmassianetal2017,PagnotaandZurek2016}. 
The exact reason for asynchronism is not known yet. Intermediate Polars (IPs), which
have a weaker field strength of about 4-30 MG \citep{2020Ferrario}, are asynchronous systems to a large extent (mostly, P$_{\rm spin}$/P$_{\rm spin}$$\sim$ 0.1)
\citep{Mukai2017, deMartinoetal2020}.  Polars show strong orbital variability at all wavelengths \citep{Schwopeetal1998}.
IPs  may be disk-fed, diskless, or in a hybrid mode in the form of disk-overflow which may be diagnosed by spin, orbital, and sideband periodicities at different wavelengths \citep{Hellier1995, Nortonetal1997, deMartinoetal2020}. The accretion flow in MCVs, close to the WD, is channeled along the magnetic field lines reaching
supersonic velocities and producing a stand-off shock above the WD surface \citep{Aizu1973}. The post-shock region is hot (kT $\sim$ 10-50  keV) and cools via thermal Bremsstrahlung producing hard X-rays 
and cyclotron radiation emerging in the optical/nIR band \citep{Mukai2017}. Both emissions are partially thermalized by the WD surface and re-emitted in the soft X-rays and /or EUV/UV domains.
The relative proportion of the two cooling mechanisms strongly depends on the field strength and the local mass accretion rate. Cyclotron radiation dominates for  high field Polars and
suppresses Bremsstrahlung cooling and high X-ray temperatures creating an optically thick soft X-ray component (Blackbody kT$\sim$ 30-50 eV) due to reprocessing 
\citep{FischerandBeuermann2001, Schwopeetal2002}. Soft X-ray emitting IPs at high accretion rates also exist \citep[see][]{EvansandHellier2007}. 
The post-shock region has been diagnosed by spectral, temporal, and spectro-polarimetric analysis
in the optical, nIR, and in X-ray regimes to have complex field topology (i.e., Polars),
and differences between the primary and secondary pole geometries and emissions \citep{Ferrarioetal2015, PotterandBuckley2018}. The complex geometry and emission properties of MCVs make these objects ideal
laboratories to study in detail the accretion processes in moderate magnetic field environments, and help understanding the role of magnetic fields in close-binary evolution. MCVs are readily detected in X-ray surveys (\xmm, \swi, \textit{INTEGRAL}) and studied since they are the brightest emitting CVs with luminosities 10$^{30-34}$ ergs~s$^{-1}$ hosting WDs with a weighted average mass of 0.77$\pm$0.02 \msun \citep{2020Shaw}. In addition, they play a crucial role in understanding Galactic X-ray binary populations \citep[see][for reviews]{deMartinoetal2020,2020Lutovinov}. 

The \spektr\ X-ray observatory was launched on July 13, 2019 from Baykonur kosmodrom \citep{Sunyaev21}. It carries two grazing incidence X-ray  telescopes -- the German built eROSITA telescope operating in the 0.2-9.0 keV band \citep{2021A&A...647A...1P} and Russian built Mikhail Pavlinsky \artxc\ telescope \citep{2021A&A...650A..42P}, sensitive in the 4.0-30.0 keV energy band with maximum sensitivity near 10~keV. The main element of the science program of SRG observatory is an all-sky survey which was planned to be carried out during a four-year timespan consisting of 8 surveys with the duration of 6 months each (i.e., two surveys each year)\citep{Sunyaev21}. An extensive program of optical identification of both \erosita\ and \artxc\ sources is under way
\citep[e.g.,][]{2020AstL...46..149K,2020AstL...46..645B,2021AstL...47..661D,2022Schwope,2022Zaznobin,2023Rodriguez}.

In this paper, we introduce a new X-ray source discovered by SRG/\artxc\ during the first sky survey in 2020 which we have identified as a Polar-type MCV. In order to identify and study the source we used observations with the optical telescopes at TUG (T\"UB\.{I}TAK National Observatory, Turkey) -- mainly RTT-150 1.5m telescope and at the SAO RAS (Special Astrophysical Observatory, Russian Academy of Sciences) -- mainly 6m telescope (BTA). We proposed and obtained follow-up X-ray observations with the \nicer\ Observatory in the low state and analyzed the existing \erosita\ data of this source in the high and low states. In the following sections, we detail our data, analyses, and results. Finally, we discuss the implications of our new findings in the context of MCVs. 

\section{The Data and Observations}\label{data:obs}

According to a protocol signed between Turkey and Russia, selected new sources discovered by SRG/\artxc\ telescope are delivered/shared with their coordinates and X-ray flux, to study optical counterparts and perform multi-wavelength observations. The X-ray source \srgcv\ was found in the first-year map of the Mikhail Pavlinsky \artxc\  telescope all-sky survey in the 4.0–12.0 keV energy band \citep{Pavlinsky22}. In the scanning mode for faint sources \artxc\  telescope  does not yield sufficient number of counts for a meaningful  spectral or temporal analyses. The source was detected  in the high state in 2020 slightly over the  5$\sigma$ threshold. These data were used to measure the 2020 flux  in the  4.0-12.0 keV energy band as (4.0-7.0)$\times$10$^{-12}$ \fluxcgs. In the year 2021, the source was found to be in a low state. \erosita\ telescope operating in the softer 0.2-9.0 keV  energy range  detected \srgcv\ with higher confidence, registering about a total of 420 photons above the background in 2020-2021 ($\sim$370 belong to the high state). Spectral analysis of these data will be presented  in  Sec.~\ref{sec:ana-xray2}. 

To study the source in more detail, we proposed and obtained X-ray data from the \nicer\ mission in the 0.2-10.0 keV range. The \nicer\ observation was performed on 2021 June 13 (obsID=4679010102) for an effective exposure of 25.3 ks yielding a source count rate of 0.273(2) c s$^{-1}$ (0.25-10.0 keV) after the background subtraction. The data were reduced and a cleaned event file was produced utilizing two different {\sc NICERDAS} software versions with a standard \nicer\ filtering and different background models -- 3C50 and SCORPEON. \nicer\ observation took place in the low state of the source (see Sec.~\ref{sec:ana-xray1} for analysis details and results).

\subsection{Optical observations}\label{data:opt}

\begin{figure}
\begin{center}
	\includegraphics[width=0.86\columnwidth]{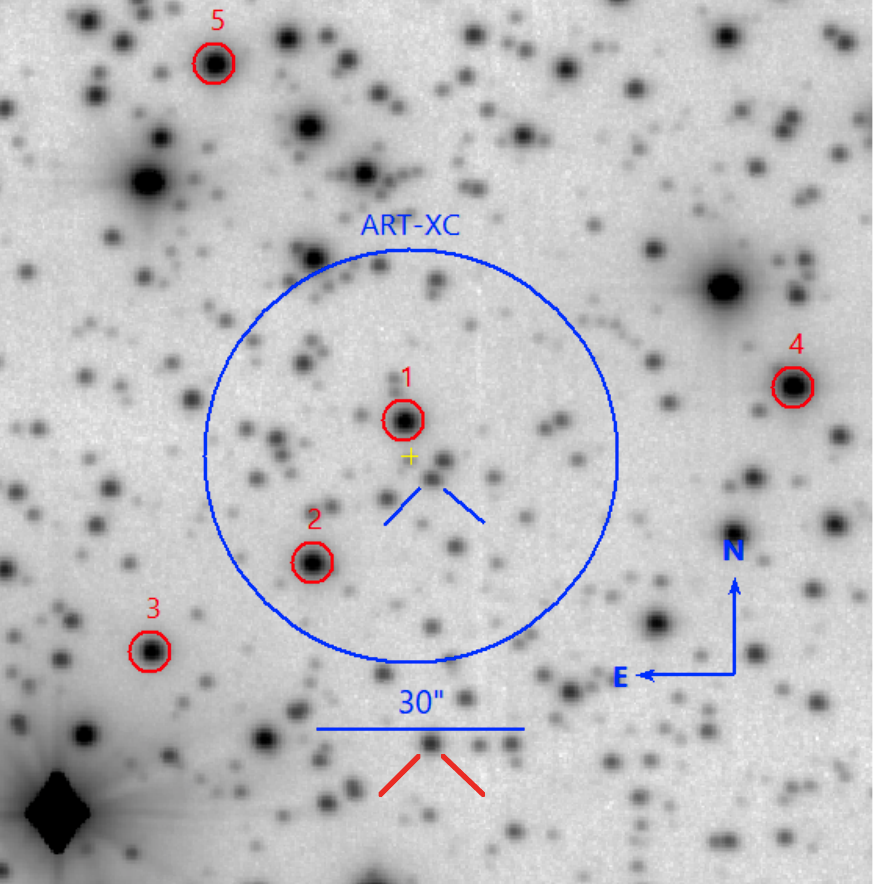}
    \caption{Optical field of \srgcv; source labeled by blue bars using \gcv\ position (also, the \erosita\ position) obtained with RTT150 using 1200 s exposure with a clear filter. Red bars show the comparison star used for CCD photometry. Red circles denote the standard reference stars in the field.}
    \end{center}
    \label{fig:field}
\end{figure}

The optical observations of the source began in 27 July 2020 at the Russian-Turkish 1.5m telescope (RTT150) using the multi-functional detector TFOSC (T\"{U}B\.{I}TAK Faint Object Spectrograph and Camera). Fig.~\ref{fig:field} shows the 1\pri.5x1\pri.5 optical image of the field around the \artxc\ new source position obtained with RTT150 using a 1200 s exposure without any filter.  The large  1\pri\ circle shown in the figure is slightly higher than the angular resolution (PSF FWHM) of the \artxc\ telescope which is  53\pripri\ \citep{Sunyaev21,2021A&A...650A..42P}. 
We selected all the sources inside the 1\pri\ error circle (shown in Fig.~\ref{fig:field}) with a limiting magnitude of 20.5 in the g-band (SDSS) suitable for spectral observations with RTT150. The total number of selected sources was 30. We applied MOS observational technique for the spectral extraction of multiple objects. Five multi-object masks were used with pinhole apertures. The description of the MOS-technique on the mask calculations with non-overlapping dispersion, and the details of the extraction, dispersion and flux calibration of individual spectra obtained from this method can be found in \citet{Khamitov20}.
Utilizing the MOS-observation technique, we found a blue source very close to the new candidate \artxc\ coordinates with strong emission lines of Balmer series and He lines. This source was the only plausible one that had the typical signatures of an interacting/accreting binary that could appear in the hard X-rays. The coordinates of the detected new source was also coincident with a \gaia\ alert source, \gcv; a suspected CV candidate \citep{2019Hodgkin}. This conclusion was confirmed by inclusion of the \erosita\ data and localization of the new source at $\alpha$=21$^h$,31$^m$,51$^s$.0\ and $\delta$=+49$^{\circ}$,14\pri,02\pripri.1  with the positional error of $\approx 5\arcsec$. Therefore, we selected this source as the highly probable optical counterpart. 

\begin{figure*}
	\includegraphics[width=17.6cm,height=10.2cm,angle=0]{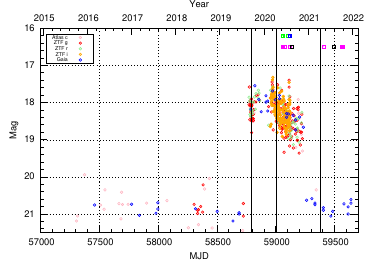}
    \caption{The archival photometry of \srgcv\ in 2015-2022. The
    observations are: 1) Gaia (blue), 2) ATLAS c (light pink), 3) ZTF g (red), r (light green) and i (yellow). The photometric and spectroscopic observations obtained from TUG (Turkey) are represented with pink and green squares, respectively while those obtained from SAO (Russia) are represented with black and blue squares. The vertical solid lines at MJD 58789, MJD 59003 and MJD 59378 indicate the dates of the \gaia\ alert, \artxc\ detection and NICER start time, respectively.}
    \label{fig:longlc}
\end{figure*}

The \gaia\ alert source, \gcv, has coordinates $\alpha$=21$^h$,31$^m$,50$^s$.80\ and $\delta$=+49$^{\circ}$,14\pri,01\pripri.68\ and it is not included in any current X-ray catalog. 
The detection history of \gcv\ is given in Table~\ref{Table:1}. The earliest \gaia\ brightness of the new source is 20.91$\pm 0.15$ mag in the G band that reveals a low state. After the alert with a flaring rise to a maximum, it remains at 17.94$\pm 0.21$ mag in the high state for about 8 months and starts declining. Our optical observations with the Turkish telescopes are made during this period when the source was bright (see Figure~\ref{fig:longlc}). The \gaia\ DR3 source ID number is 1978752971571052672, there is no calculated parallax thus no distance estimate is possible. 

Figure~\ref{fig:longlc} shows the long-term light curve of the new source covering the period from 2015--2022 constructed using the existing \gaia\ data together with the light curves from  archival optical photometric surveys. We were able to  retrieve the observational data  from ZTF\footnote{The ZTF data can be obtained from IRSA https://irsa.ipac.caltech.edu/Missions/ztf.html. The 5-sigma limiting magnitude per exposure is $\sim 20.5$ mag.} (Zwicky Transient Facility; \textit{g}, \textit{r} and \textit{i} filters) \citep{Masci19} between August 2018 and January 2021,  and ATLAS\footnote{The ATLAS Forced Photometry is available at https://fallingstar-data.com/forcedphot/. The ATLAS sky survey telescope observes the available sky every few nights in two filters: cyan (420–650 nm) and orange (560–820 nm). The approximate 5-sigma limiting magnitude per exposure is 19.7 for both filters.} (Asteroid Terrestrial- impact Last Alert System; \textit{o} and \textit{c} filters) \citep{Shappee14, Jay19}  between July 2015 and November 2021. For ATLAS, we removed poorer quality data where the uncertainty in the observed magnitude was $>$ 2.5 magnitudes. The o-band data of ATLAS are more scattered than the c-band and other survey data with large uncertainty in the magnitudes, thus,  we did not use it for the plot.  The total  long-term light curve indicates a low accretion state ($\sim 21$ mag) that was detected around 4-4.5 yrs with ATLAS and \gaia\ which switched to a high state ($< 18-19$ mag) that lasted slightly over a year.  In addition to Gaia and ATLAS observations, ZTF data support that the system was in a higher state between November 2019 and January 2021, $ \sim 12$ months.  
We note that  \citet{2021Szkody} present also ZTF data  covering partly the year 2020 for \srgcv\ with an optical spectrum obtained using the Keck Telescope suggesting  the source is a MCV candidate.

In addition to RTT150 photometry, optical photometric observations of the optical counterpart are  conducted with other Turkish and Russian telescopes (1-m T100 at TUG and 1-m Zeiss-1000 at SAO RAS) and available smaller telescopes  ADYU60 (60 cm) in Ad{\i}yaman University Observatory. A log of all photometric observations from Turkey and Russia is given in Table~\ref{Table:3}. Following the identification of the optical candidate from RTT150 photometry and MOS spectroscopy as described in the previous paragraphs, more spectral observations were performed in the standard mode with long slit aperture using RTT150 at TUG followed by spectroscopic observation using BTA at SAO RAS. A detailed list of spectroscopic observations is displayed in Table~\ref{Table:4}.

\begin{table}
\caption{Observational timeline of the \srgcv\ X-ray source} 
\label{Table:1} 
\centering 
\begin{tabular}{l l l}
\hline\hline 
Observational Detection                    & Date           & Time           \\
                                           & (YY.MM.DD)     & (HH:MM:SS)     \\
\hline 
Gaia alert                                 & 2019.11.02     & 08:05:26 (UTC) \\
Detection - SRG/\artxc\                  & 2020.06.02     & 23:27:12 (UTC) \\
                                           & 2020.06.03     & 23:18:49 (UTC) \\
Discovery - RTT150                      & 2020.07.27     & 22:32:43 (UT)  \\
\hline 
\end{tabular}
\end{table}

\begin{table*}
\caption{A log of photometric observations of \srgcv\ } 
\label{Table:3} 
\centering 
\begin{tabular}{c c c c c c c c}
\hline\hline 
Date      & HJD Start      & Duration & Exposure & Number of Frames & Filter & Telescope    & Observatory \\
(YY.MM.DD)           & (HJD 2459000+) & (h)      & (s)      &                  &        &              &             \\
\hline 
2020.07.28             & 059.24174      & 7.1      &  60      & 280              & clear  & RTT150       & TUG \\
2020.07.30             & 061.27805      & 1.2      &   7      & 315              & clear  & RTT150       & TUG \\
2020.08.01             & 063.27149      & 2.0      &   7      & 614              & clear  & RTT150       & TUG  \\
2020.08.02             & 064.28747      & 2.2      &  30      & 224              & clear  & RTT150       & TUG  \\
2020.08.03             & 065.37281      & 4.6      &  30      & 474              & clear  & RTT150       & TUG  \\
2020.08.17             & 079.34191      & 5.6      &  90      & 208              & V      & RTT150       & TUG  \\
2020.08.27             & 089.40084      & 3.9      & 240      & 44              & clear  & ADYU60       & ADYU-ARMER  \\
2020.08.28             & 090.24964      & 8.3      & 240      & 107              & clear  & ADYU60       & ADYU-ARMER  \\
2020.09.10             & 103.22123      & 8.1      & 240      & 120              & clear  & ADYU60       & ADYU-ARMER  \\
2020.09.14             & 107.23328      & 8.9      & 240      & 118              & clear  & ADYU60       & ADYU-ARMER  \\
2020.09.15             & 108.33785      & 5.1      & 240      & 77              & clear  & ADYU60       & ADYU-ARMER  \\
2020.09.28             & 121.34784      & 3.7      & 300      & 38              & g      & RTT150       & TUG  \\
2020.10.13             & 136.15548      & 3.1      &  120     & 72              & clear       & Zeiss-1000 & SAO \\
2020.10.16             & 139.31377      & 2.3      &  10      & 579              & clear  & T100         & TUG   \\
\hline
2021.07.12 & 408.45278 & 1.6 & 300 & 16 & g,r,i & RTT150 & TUG \\ 
2021.10.09 & 497.18620 & 2.7 & 300 & 31 & clear & Zeiss-1000 & SAO \\ 
2021.12.09 & 558.19185 & 2.7 & 300 & 28 & g & RTT150 & TUG\\
2021.12.17 & 566.18888 & 0.1 & 300 & 1 & g & RTT150 & TUG\\
2021.12.28 & 577.19960 & 0.1 & 300 & 1 & g & RTT150 & TUG \\
\hline 
\end{tabular}
\end{table*}

\begin{figure*}
	\includegraphics[width=\textwidth]{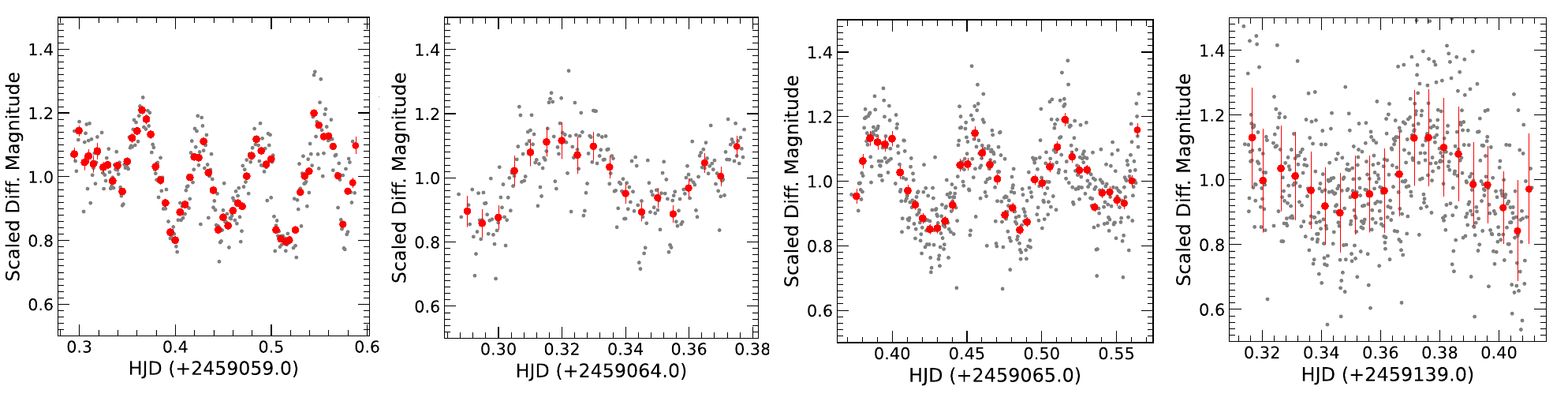}
\vspace{-0.5cm}
	    \caption{
     A sample of \srgcv\ light curves obtained at TUG covering $\geq 2$ hours duration (with clear filter). Observation dates and telescopes from left to right are: 2020-07-28 (RTT150), 2020-08-02(RTT150), 2020-08-03 (RTT150), 2020-10-16 (T100). Black circles are unbinned and red circles represent data binned at 7.2 min.}
    \label{fig:lcs}
\end{figure*}

\section{Analysis and Results}\label{analysis-results}

\subsection{Optical photometry using ground-based facilities during the high state in 2020}

We performed photometric observations of \srgcv\ between July and October 2020 (during its high state) with three different telescopes in Turkish sites, as follows: Ritchey-Chretien type 1.5-m (TUG RTT150),
1-m (TUG T100) telescopes at the TUBITAK National Observatory (TUG) equipped with Andor iKon-L 936 BEX2-DD-9ZQ  and SI1100 CCD cameras,  and 0.6-m telescope at the Adiyaman University Application and Research Center (ADYU60) equipped with 1k $\times$ 1k Andor iKon-M 934 CCD with a pixel scale of 13$\mu$m $\times$ 13$\mu$m. 
Details like telescopes, exposure times, and filters are listed in Table~\ref{Table:3}. 

Standard CCD reductions were performed on the data obtained from the  2020 observations. Since the dark counts were negligible for the used CCDs, only bias subtraction and flat-field corrections were applied.  To extract instrumental magnitude of the sources in the field, we used aperture photometry through a developed script using {\tt Python} and {\tt Sextractor}. We selected a comparison star with similar brightness to the new optical counterpart, \gcv, in the field of view. The comparison star is shown in Fig.~\ref{fig:field} with red bars. The \gaia\ source ID is 1978752975879142528 (DR3) with $\alpha$=21$^h$,31$^m$,50$^s$.82\ and $\delta$=+49$^{\circ}$,13\pri,23\pripri.26\ ($g$=17.64). We chose a small aperture $\sim1.2\times$FWHM since there is another source very close to \gcv. Because our source is quite faint, relatively high SNR (signal-to-noise ratio) values could not be obtained. For each night, a light curve was constructed using instrumental magnitude differences
between the comparison and our optical counterpart.  We subtracted the nightly mean differential magnitudes and converted times to HJD using {\tt Astropy} timing analysis package \footnote{http://www.astropy.org}.  
Example light curves which have adequate number of data points covering $\geq2$ hrs time-span  are given in Fig.~\ref{fig:lcs}. 

\begin{figure}
 \vspace{-0.4cm}
\hspace{-0.3cm}
	\includegraphics[width=9cm,height=7.5cm,angle=0.0]{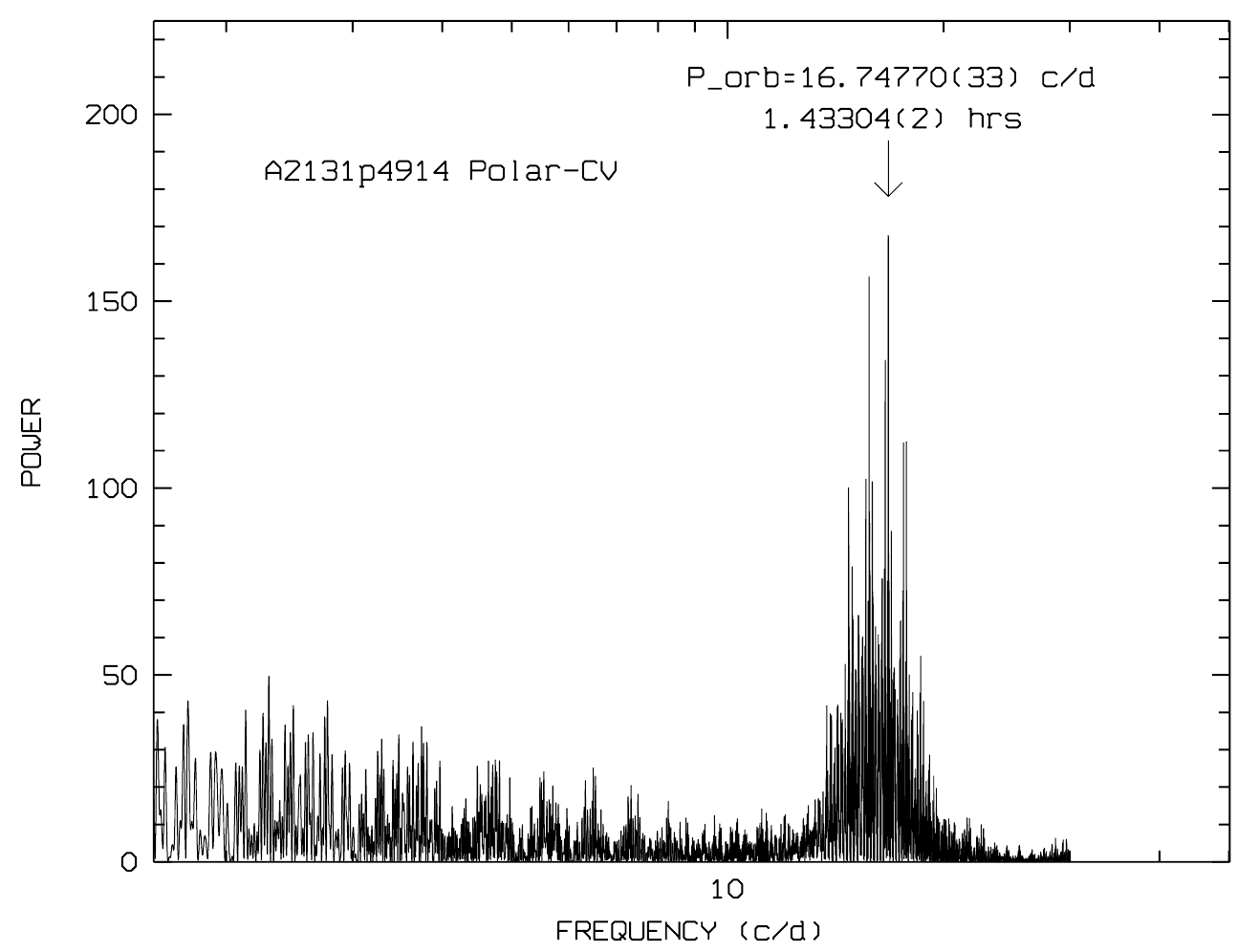}
 
 \vspace{-0.7cm}
 
 	\hspace{-0.35cm}
		\includegraphics[width=10cm,height=6.3cm,angle=0.0]{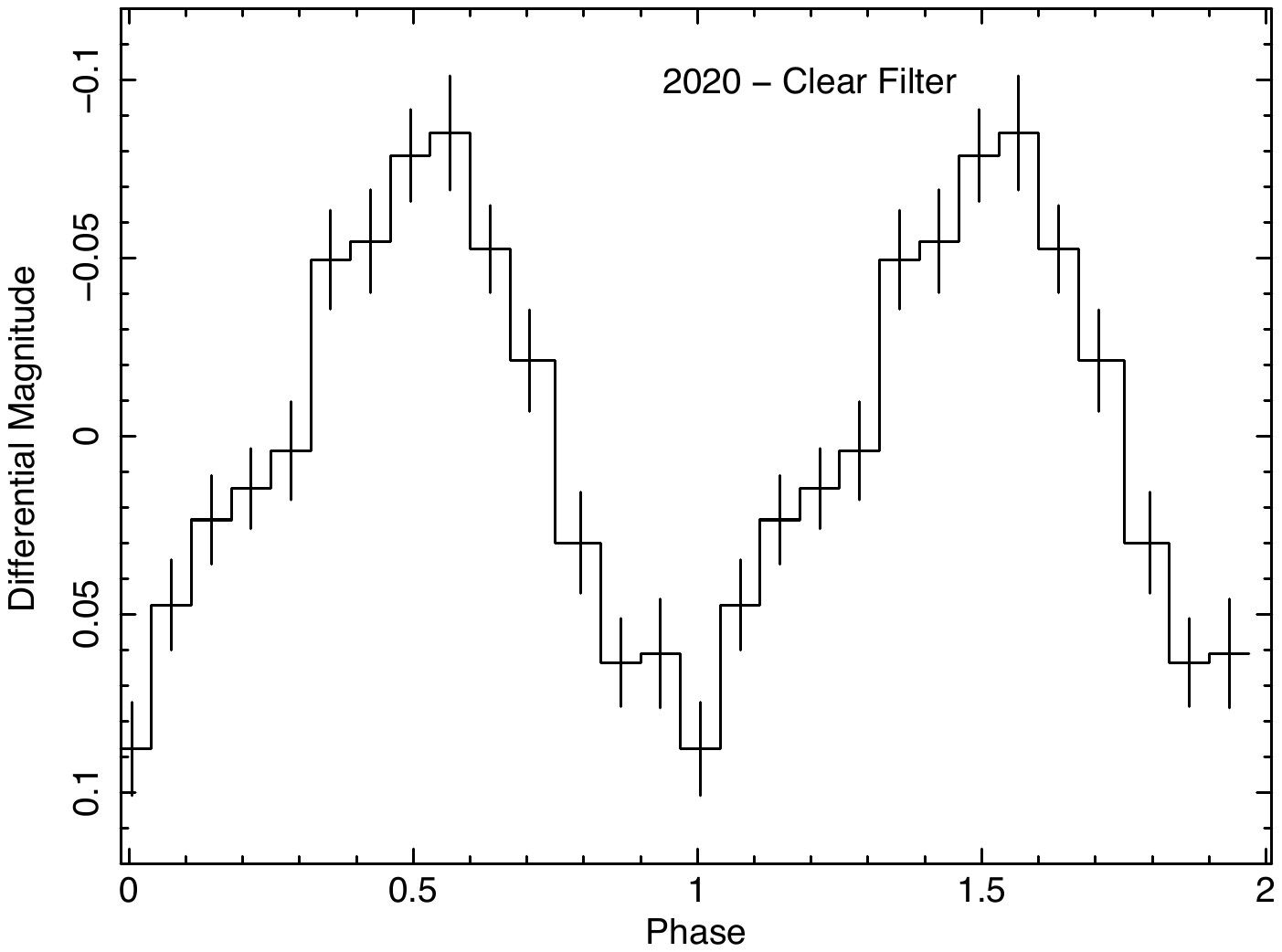}
	 \vspace{-0.6cm}
    \caption{Top panel is the power spectrum of \srgcv\ derived from the entire ground-based photometric light curve in July-October 2020 (high state). The detected period is labeled on the figure. The bottom panel shows the mean light curve derived from the same data set in 2020 during the high state. The phases on the x-axis are with respect to the T$_{min}$ of the Ephemerides.}
    \label{fig:PDSbalman}
\end{figure}

\subsubsection{Power spectral analysis of the photometric light curves obtained at TUG and  ADYU60 during the high state}

The resulting light curves created from (differential) aperture photometry are used for power spectral analysis. To obtain the periodicity on individual nights, power density spectra (PDS) were constructed using Lomb-Scargle algorithm \citep{1976Ap&SS..39..447L,1982Scar, 2018ApJS..236...16V}, provided by {\tt Astropy} \citep{2013A&A...558A..33A,2018AJ....156..123A}. 
The PDS for individual nights show significant single peaks above $3\sigma$ significance for a periodicity between 82 and 93 minutes.

Furthermore, we performed power spectral analysis using the entire photometric light curve of the RTT150 and ADYU60 data obtained with the clear filter using the TSA (time series analysis) context within MIDAS software package version  17FEBpl1.2\footnote{https://www.eso.org/sci/software/esomidas/news/17feb\_pl1\_2.html} and the Lomb-Scargle algorithm. In order to correct for the effects of windowing and sampling functions on power spectra, synthetic constant light curves are created and a few prominent frequency peaks that appear in these light curves are pre-whitened from the data in the analysis. Before calculating the total PDS, the individual or consecutive nights are normalized by subtraction of the mean magnitude. Moreover, when necessary, the red noise in the lower frequencies is removed by detrending the data. Fig.~\ref{fig:PDSbalman} top panel shows the final PDS constructed from the total photometric data using 1260 frames (for 2020 data). There is only one very prominent peak well above 3$\sigma$ significance \citep[see][]{1982Scar} at  16.74770(33) c/d corresponding to 1.43304(2) hrs or 85.982 min as marked on the figure and the other peaks around the period are 1d-2d aliases. The error on the detected frequency is the half-width of the power spectral line at 99.99\% enclosed power. This is calculated using the {\tt width/tsa} task within the TSA context of MIDAS. The bottom panel of the same figure shows the folded mean optical light curve obtained using our time series in 2020 with the clear filter. The semi-amplitude of variations is around 0.3 mag and decreases to 0.1 mag once pre-whitening and detrending is applied. Using the same data set we derived Ephemerides for the minimum times of this period utilizing a securely recovered modulation dip via fitting a sine curve as

\begin{equation}
{\rm T_{min} = HJD\ 2459059.3661(3) + 0.059710(1)E}
\label{ephemeris}
\end{equation}

The detected period is a typical orbital period of a CV below the period gap (the gap is at 2-3 hrs). On the other hand, since we have not found any other optical periodicity,  we have to rely on the spectroscopic and X-ray data analysis for further classification of the CV candidate. As far as the photometry is concerned this may be an MCV most likely a Polar-type since we do not find a separate spin period. It may also be a nonmagnetic CV below the period gap, however the significant high and low state transitions are atypical of this class and more typical of polar-type MCVs. 

\begin{figure}
\hspace{-0.35cm}
 \includegraphics[width=9cm,height=5.5cm,angle=0.0]{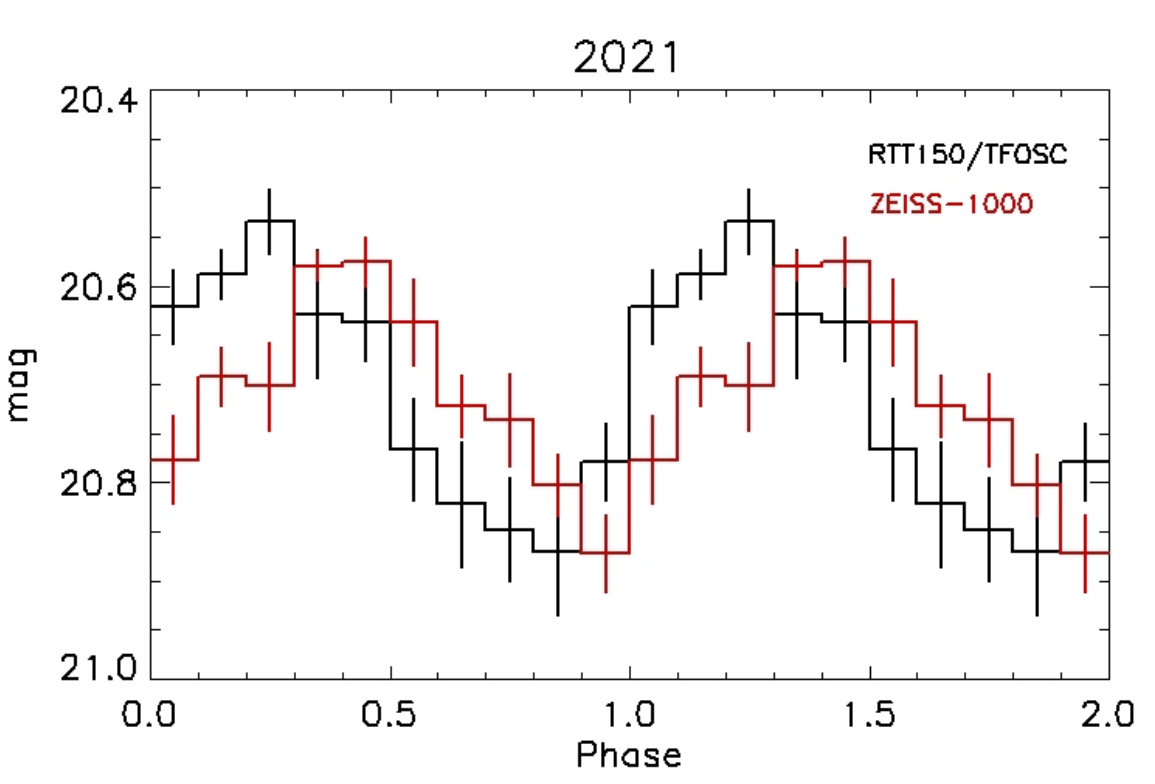}
    \caption{The mean light curve of \srgcv\ obtained with RTT150 in December 2021 using the $g$-band filter (SDSS) during the low state (in Black). Overplotted is the mean light curve of the source obtained with Zeiss-1000 in 09 October 2021 using a clear filter (in red).\label{fig:lc2021} }
\end{figure}

\subsection{Follow-up photometry using TUG and SAO during low state in 2021}

We only performed occasional follow-up photometric observations of \srgcv\ in 2021 with SDSS filters since the source was very dim for time series analysis with RTT150. Table~\ref{Table:3} shows the log of observations in 2021 during the low state.
PSF photometry is performed for the analysis. Fig.~\ref{fig:lc2021}\ displays the $g$-band (SDSS) photometry and the mean photometric light curve (in black) of the December 2021 RTT150 data phase-locked to our derived photometric Ephemerides. The pulse profile resembles the profile in the high state but with about 50\% decrease in semi-amplitude at $\sim$ 0.15 mag. The accumulated phase error for December 2021 data is about 0.16~ and the photometric peak is at 0.2-0.3 phase (note that this is not obtained with clear filter). Additional photometric observations were obtained at 1m Zeiss-1000 Telescope of SAO RAS (see
Table~\ref{Table:3}). Observations were carried out both in high (2020 Oct) and low (2021 Oct) brightness states of \srgcv. The obtained data were processed in a standard way using the {\sc IRAF}\footnote{Developed by USA National Optical Astronomical Observatories (NOAO)-\\https://iraf-community.github.io/} package tools. PSF photometry was carried out. We have overplotted the mean light curve obtained from Zeiss-1000 in October 2021 (low state) using a clear filter over the December 2021 mean light curve in Fig.~\ref{fig:lc2021}. The photometric pulse peaks at about 0.4 phase for October 2021 consistent with the peak phase in the high state within the accumulated phase error of 0.13\ . The slight changes in the pulse and the phase of the peak may occur as a result of the error in the period.  On the other hand, the changes of accretion geometry (in high and low states), and the changes in hot spot morphology, together with usage of different filters which may reflect real color variations, can result in such phase shifts at maximum pulse-phase.

\subsection{Optical spectroscopy using ground-based facilities during the high state in 2020}

\subsubsection{Long-slit spectroscopy with RTT150 at TUG}

Spectroscopic data, taken in approximately five weeks period, correspond to the high state of \srgcv\ starting about two months after \artxc\ discovery. A log of RTT150 spectroscopic observations are given in Table~\ref{Table:4}. For convenience, the times of the observations are labeled in Fig.~\ref{fig:longlc} over the  \gaia\ light curve. Long-Slit (LS) spectroscopy was performed with three different grisms;  No.7 (4220-6650 \AA), No.8 (6190-8190 \AA), and the broad-band No.15 (3650-8740 \AA). The slit aperture is 1\pripri.78. The resolution capacities of the grisms are: 1) No.15 ($\lambda/\Delta\lambda$)=749, 2) No.7  ($\lambda/\Delta\lambda$)=1331, and 3) No.8 ($\lambda/\Delta\lambda$)=2189. The exposure times for\ LS spectroscopy are 900 s and 1200 s for grisms No.15 and No.7-8, respectively.  Spectrophotometric standard stars from \citet{Oke1974} and \citet{Stone1977} catalogs were observed each night. Biases, halogen lamp flats, and Fe-Ar calibration lamp exposures were obtained for each observation set. Preliminary reduction steps were completed using task {\tt ccdproc} within {\sc IRAF}. The reductions and analysis of spectra were made using the Long-slit context of {\sc IRAF} software. 

\begin{figure*}
	\includegraphics[width=\textwidth,height=12cm]{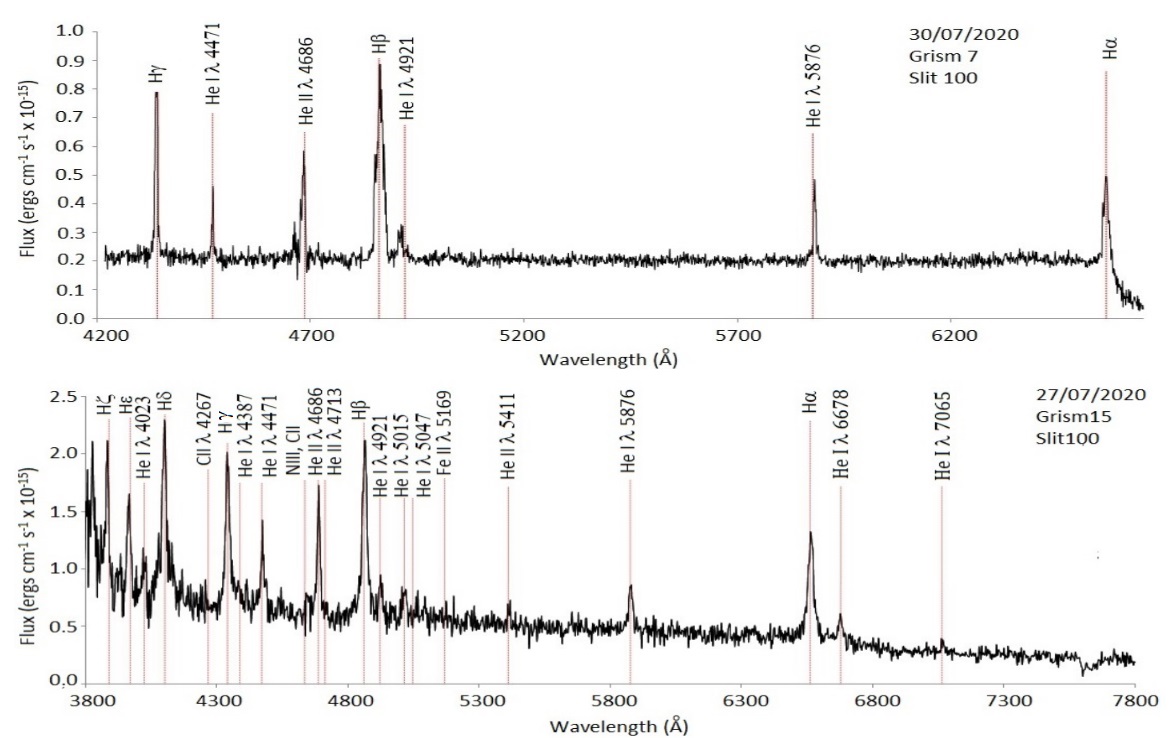}
    \caption{Two Long-slit spectra of \srgcv\ obtained during the high state with grisms No.7 and 15 in the top and bottom panels, respectively. Dates of observations are labeled on the panels. Some emission lines like Balmer lines, low ionisation lines like He I and Fe II are marked with red on the spectra.\label{fig:figure6}}
\end{figure*}

Fig.~\ref{fig:figure6} shows the reduced, calibrated and fluxed spectra (for convenience two different grism spectra are displayed). Some of the identified lines are labeled on the figure. LS spectrum obtained in July 2020 with grism No. 15 has been used for line identification since it has the widest wavelength band with a modest spectral resolution. We list the brightest lines (with S/N $>$ 3.0) and relative fluxes (unabsorbed) in Table~\ref{linelist}.  Line identification and flux determination are performed using the {\tt splot} task and Gaussian fitting within the {\sc IRAF} software. 
We find that the spectrum of \srgcv\ increases towards the blue continuum with strong hydrogen and helium emission lines revealing single-peaked and narrow structures. Balmer lines are the strongest. We find He\,{\sc ii} ($\lambda$4686) to H$\beta$ ratio as 0.74 (see Table~\ref{linelist}). This indicates existence of highly ionized medium via irradiation and reprocessing. Such strong ionized helium lines are typical of MCVs with a criterion of He\,{\sc II}/H$\beta$ $\ge$ 0.4 (higher for polars) for classification \citep[][]{1992Silber} which points out that our source belongs to this class of variables. None of our spectra in the high state showed obvious cyclotron humps,  however further spectral modeling with WD, secondary star, and disk emission models to investigate the existence of cyclotron excess may improve this. 

\begin{figure}
\vspace{-0.6cm}
\hspace{-0.4cm}
	\includegraphics[width=9.9cm,height=6.5cm]{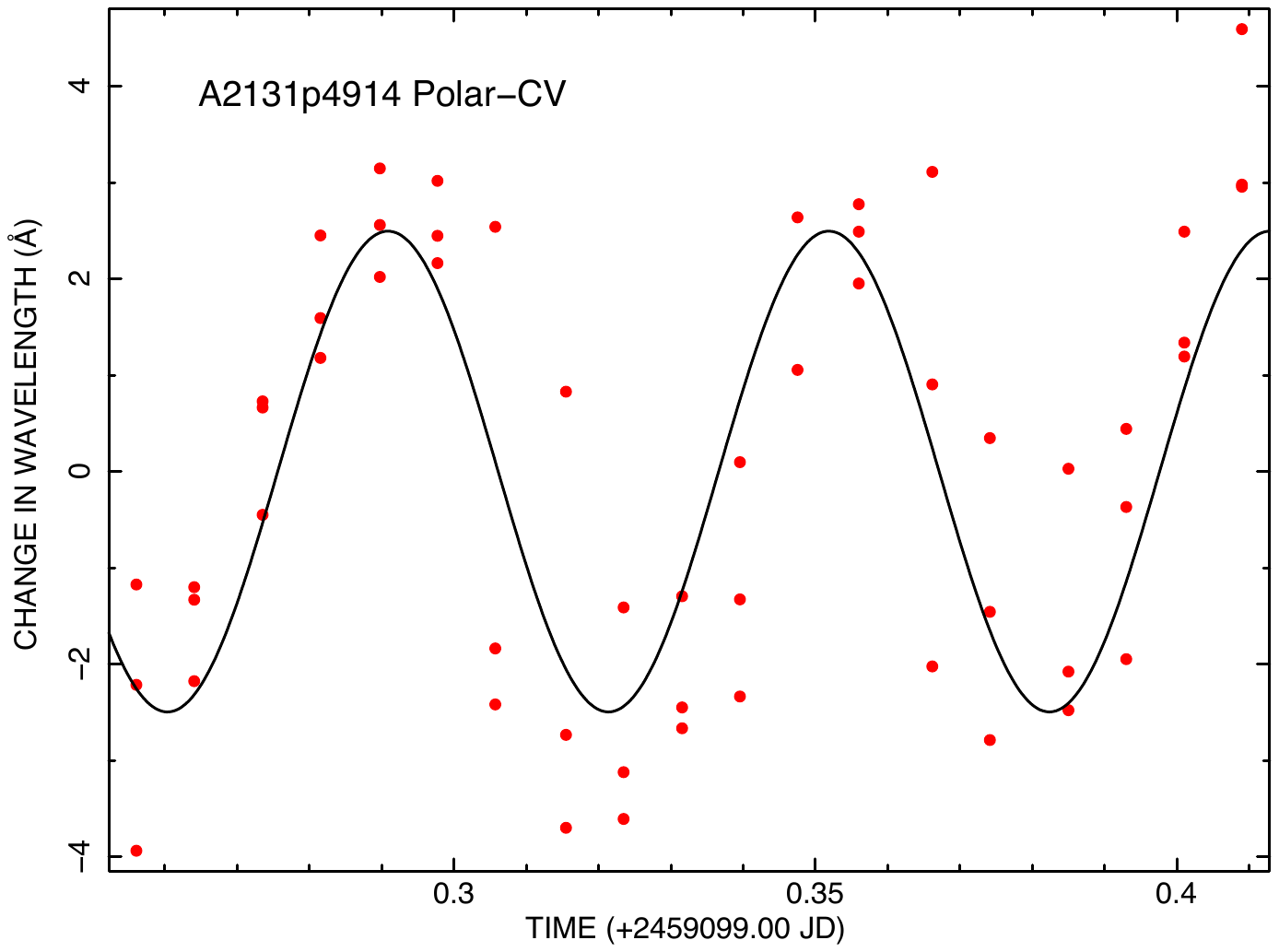} 
    \caption{The variations of relative change in the wavelength of the H$\gamma$, H$\beta$, and H$\alpha$ lines versus time covering three periods.\label{fig:spec-periBalman}}
\end{figure}

\subsubsection{Phase-resolved MOS spectroscopy with RTT150 at TUG}

We obtained 19 consecutive optical spectra each 600 s using MOS spectroscopy (see Sec.~\ref{data:obs}\ for details). In contrast to the masks with pinhole apertures used to identify the source, the short slits (SS) were used in this mask with non-overlapping dispersion. In addition to the \srgcv\, SS were also used at the positions of secondary standards for differential spectrophotometric calibration and control of the position of the source on the slit for more correct dispersion calibration.  The total duration of the MOS-spectral observations was about 3.5 hrs covering about three orbital/spin periods.

The MOS spectra are analyzed in the same manner with the LS spectra using the {\sc IRAF} software. A reduced and flux-calibrated set of spectra are calculated without sky subtraction to yield good S/N around prominent lines and calculate the variations of line centroids while following sky lines for calibration purposes. We selected three well recovered Balmer lines H$\gamma$, H$\beta$, and H$\alpha$ and derived the line centroids via fitting Lorentzian profiles along with a constant continuum value around the line of interest. Fig.~\ref{fig:spec-periBalman} shows the calibrated and normalized variation of all three line centroids where the mean value of the detected line center is subtracted. We do not find any phase-shifts between different Balmer line variations (see $\phi_0$ in \eqref{eqn:rv}) which help to construct the figure, but there are different amplitude variations for each line center that mimic y-error. We followed a sky line of O\,{\sc i} at 5577 \AA\ (green line) for calibration purposes of line centroids which showed a variation $< |0.3|$ \AA\ all throughout the 3.5 hrs, considerably less than the centroid variations of Balmer lines. Next, we analyzed the data in Fig.~\ref{fig:spec-periBalman}  using Lomb-Scargle algorithm that yielded a spectroscopic period of  0.0609(20) d (detected well over 3$\sigma$ significance). The error corresponds to the half-width of the power spectral line at 99.89\% enclosed power. We checked the individual line variations and they all show a similar periodicity. This period is longer than the photometric period we calculated by about 2\%. However, the error range of the spectroscopic period overlaps with the photometric period making this a regular polar-type MCV system. 

\begin{table*}
\caption{ Spectroscopic observations of \srgcv\ obtained at TUG, Turkey and SAO, Russia.} 
\label{Table:4} 
\centering 
\begin{tabular}{r c c c c c c c c}
\hline\hline 
Date             & HJD Start      & Duration & Exposure  & Number of Frames & Grism                 & Telescope    & Observatory \\
                 & (HJD-2459000+) & (h)      &  (s)      &                  &                       &              &             \\
\hline 
20200727         & 058.35221      & 1.5      & 1800      &   3              & 15                    & RTT150-MOS   & TUG \\
0727             & 058.50722      & 0.8      &  900      &   3              & 15                    & RTT150       & TUG \\
0727             & 058.55985      & 0.5      & 1800      &   1              & 15                    & RTT150-MOS   & TUG \\

0730             & 061.47788      & 1.2      & 1200      &   3              &  7                    & RTT150       & TUG \\
0730             & 061.53521      & 0.7      & 1200      &   2              &  8                    & RTT150       & TUG \\
0906             & 099.16971      & 3.5      &  600      &  19              &  7                    & RTT150-MOS   & TUG \\
0925             & 118.22226   &  1.8    &   300    &   20   &  VPHG1200G           &  BTA        & SAO RAS\\
 0926           &  119.30117   &  0.2     &   300     & 3               &  VPHG550G             & BTA        &  SAO RAS \\
\hline 
\end{tabular}
\end{table*}

\begin{table}
\small
\begin{center}
  \caption{Emission line fluxes (absorbed) from RTT150 in July 2020 }\label{linelist}
\begin{tabular}{llll}
\hline
Feature& $\lambda$ &Flux &S/N \\
&({\r{A}}) &  10$^{-14}$ erg s$^{-1}$ cm$^{-2}$ {\r{A}}$^{-1}$& \\
\hline
H$\delta$ &  4102 & 2.2&5.8 \\
H$\gamma$ & 4341 & 2.9&9.5\\
He\,{\sc I}&  4471& 1.4&7.4\\
He\,{\sc II}&  4686& 2.8&10.4\\
H$\beta$ & 4861 & 3.8 &14.2\\
He\,{\sc I}&  4922& 0.7 &3.4\\
He\,{\sc I} & 5876& 0.75&4.3\\
H$\alpha$ &6563& 2.45&8.8\\
He\,{\sc I} & 6678& 0.35&3.3\\

\hline
\end{tabular}
\end{center}
\end{table}

\subsubsection{Long-slit spectroscopy with BTA at SAO, RAS and the radial velocity curve}

\begin{figure}
	\includegraphics[width=0.88\columnwidth]{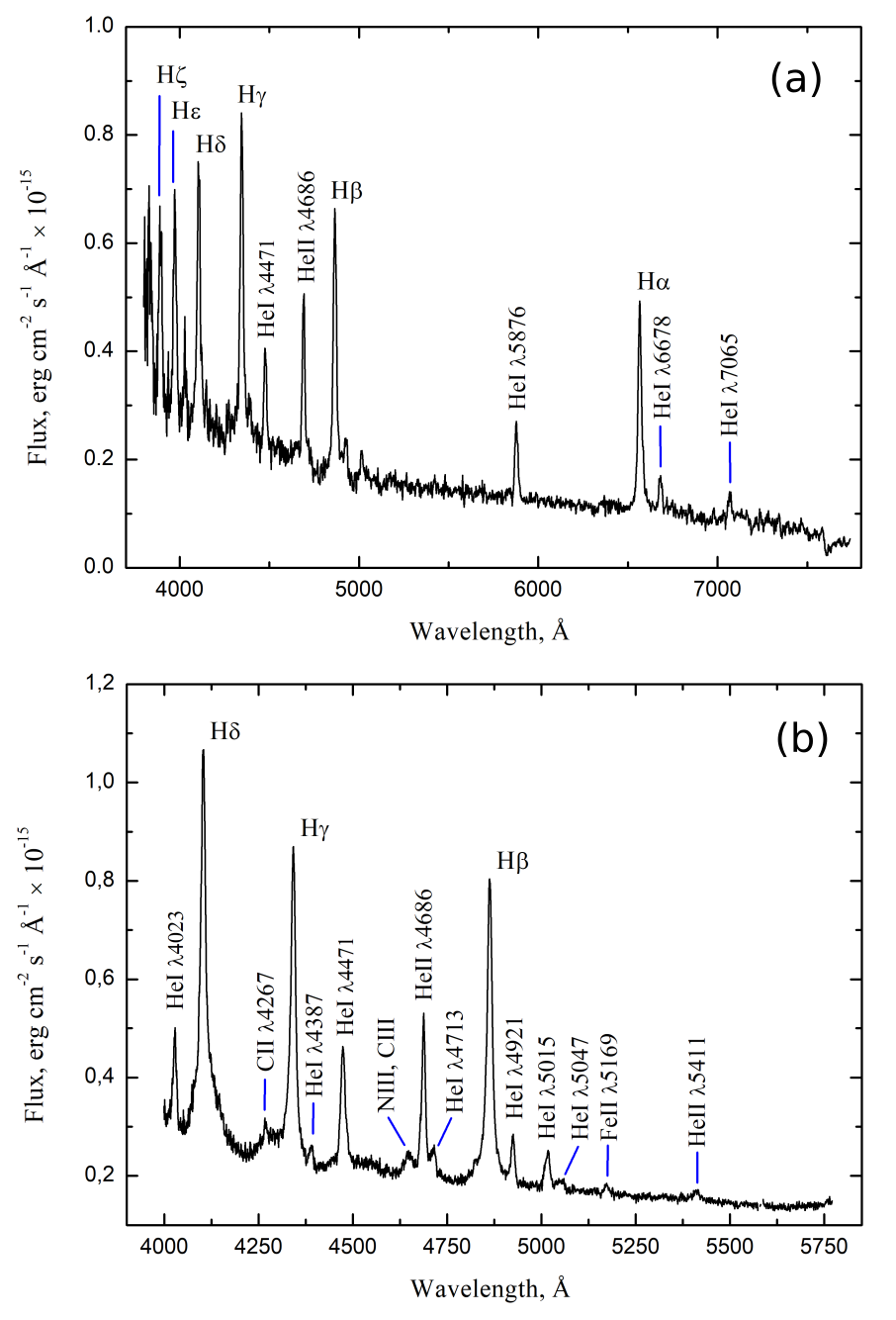}
    \caption{The average BTA spectra obtained by VPHG550G (a) and VPHG1200G grisms (b).}
    \label{fig:specs_bta}
\end{figure}

Since \srgcv\ is a dim source for RTT150, particularly for time-resolved studies,
additional spectral observations of \srgcv\ were carried out at 6m telescope BTA of Special astrophysical observatory (SAO) of Russian academy of sciences. The telescope is equipped with the SCORPIO-1 focal reducer in long-slit spectroscopy mode \citep{afan11}. The observations were performed at nights Sep 25, 2020 and Sep 26, 2020 in good astroclimatic conditions with seeing 1\pripri.2-1\pripri.5. These dates are about four months after the \artxc\ discovery during the onset of decline from the high state. At the first night, the grism VPHG1200G was used which provides an effective spectral resolution $\sim$2.6 \AA/pix with 1\pripri.2 slit width covering 3900-5800~\AA~spectral region. 20 spectra were obtained in 2 hours.  In the second night three spectra were obtained using VPHG550G grism with corresponding effective spectral resolution $\sim$6~\AA/pix in the 3500-7500~\AA~spectral region. All spectra were obtained with an exposure of 300 s.
The data reduction was carried out using {\sc IRAF} package in a standard fashion. Cosmic ray events were subtracted using {\it LaCosmic} algorithm based on Laplacian edge detection technique \citep{dokkum01}. Pixel-to-pixel variations were removed by flat field exposures. The He-Ne-Ar arc-frames were used for wavelength calibration and geometrical corrections. The spectra were extracted by optimal extraction technique \citep{horne86} with the subtraction of the background light. The flux calibration was performed using the spectra of standard star G191B2B. A signal to noise ratio of $S/N=7-8$ was acquired in the 4300-5000~\AA~range with the VPHG1200G grism during the entire observation excluding the last three spectra which were obtained under light cloud cover.

The average BTA spectra obtained with VPHG550G and VPHG1200G grisms are shown in Fig. ~\ref{fig:specs_bta}a and ~\ref{fig:specs_bta}b, respectively. Note that before averaging, the spectra were shifted to the same radial velocity $RV\approx 0$~km/s. The presented spectra are typical for magnetic cataclysmic variables. The main features of the RTT150 LS spectra are detected in the BTA LS spectra as well. 
There is strong He\,{\sc ii}$\sim$ $\lambda$4686 line of the strength $\approx \frac{1}{2}$  of H$\beta$ typical for MCVs. The VPHG550G spectra don’t show any “humps” which could be interpreted as cyclotron harmonics.  We note that the BTA LS spectra may not be very similar  in flux (line or continuum) to the RTT150 LS spectra of TUG since the source brightness had started to diminish in September 2020 and there are about two months between RTT150 and BTA spectra taken at the end of September 2020.

\begin{figure}
\vspace{-0.3cm}
\hspace{-0.3cm}
	\includegraphics[width=9.2cm,height=7.0cm]{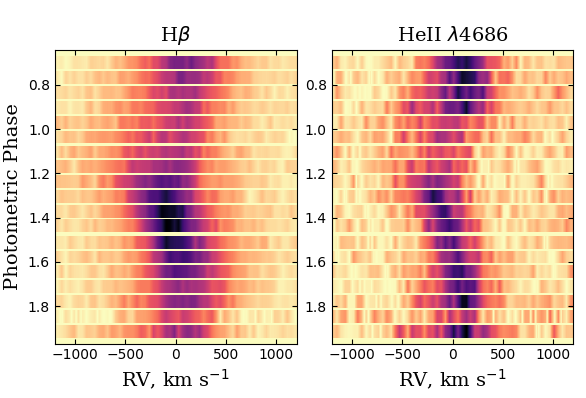}
    \caption{The dynamical spectra of H$\beta$ and He {\sc ii}~$\lambda$4686 lines.}
    \label{fig:dynam_spectra}
\end{figure}

Dynamical spectra of H$\beta$ and He\,{\sc ii}~$\lambda4686$ lines are shown in Fig. ~\ref{fig:dynam_spectra}. The figure shows the evolution of the spectral profiles during the observations. The photometric phases of spectral observations are calculated using the Ephemerides (\ref{ephemeris}) and  plotted along the Y-axis. It is seen that the lines exhibit variability due to radial velocity and intensity changes. There are noticeable differences in the behavior of He\,{\sc ii} $\lambda$ 4686 line in comparison with the H$\beta$ line. All lines are single-peaked at all observed phases, indicating the absence of an optically thin accretion disk, for which double-peaked line profiles would be expected. This agrees well with our assumption of the polar-type MCV nature of \srgcv and consistent with the results of RTT150 TFOSC spectra. 

\begin{figure}
\hspace{-0.3cm}
	\includegraphics[width=9.2cm,height=5.85cm]{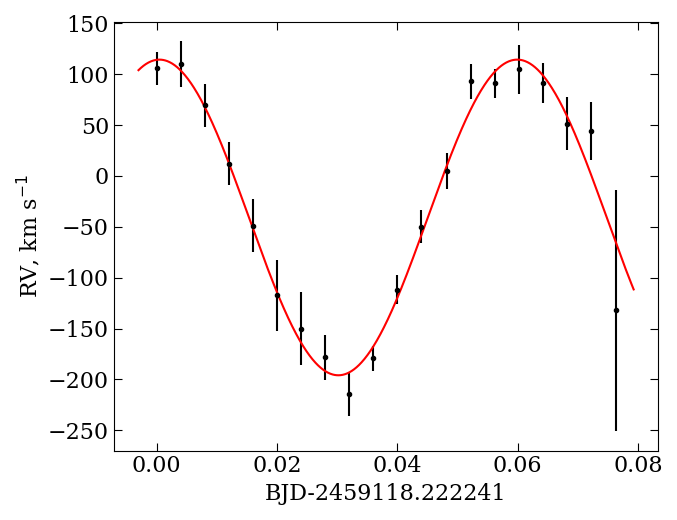}
    \caption{The observed radial velocity curve fitted by the sine function using the detected spectral period.}
    \label{fig:rvs}
\end{figure}

The radial velocities of emission lines were calculated by cross-correlation technique of \cite{tonry79} applied in the 4000-5500~\AA~region of the VPHG1200G spectra. The determination of radial velocities was performed in two steps. At the first step, the radial velocities were determined using the first spectrum as a template.  At the second step, the derived velocities were removed from each spectrum and the spectra with new wavelength scales were averaged. Then, we have improved radial velocities using the average spectrum as a template. The resulting radial velocity curve is shown in Fig.~\ref{fig:rvs}. 

\begin{figure}
	\includegraphics[width=\columnwidth]{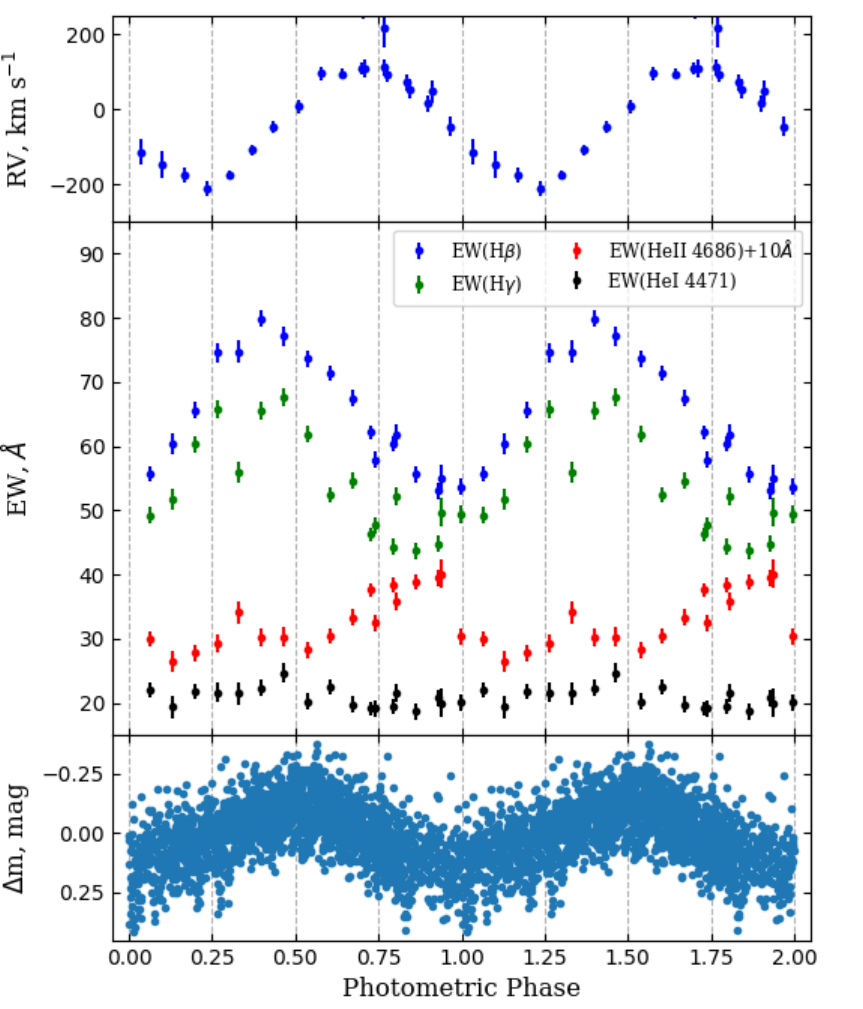}
    \vspace{-0.4cm}	
    \caption{Radial velocity curve (upper panel), equivalent width curves of four emission lines (middle panel) and photometric light curve (lower panel) of \srgcv. All curves were phased according to Ephemerides (\ref{ephemeris}).}
    \label{fig:ews}
\end{figure}

The spectroscopic period was determined by the least square fitting of the observed RV-curve by the function
\begin{equation}
    V_r = \gamma + K \sin \Big( 2\pi \frac{t}{P} + \phi_0 \Big),
    \label{eqn:rv}
\end{equation}
where $P$ is the desired period, $t$ is the epoch of mid-exposure, $K$ is the radial velocity semi-amplitude and  $\varphi_0$ is phase shift. The best-fit orbital period is $P=85.6 \pm 2.2$ min ($\chi^2_{\nu} = 0.6$). Uncertainty of the orbital period has been found by Monte-Carlo error back-propagation technique assuming normal distribution of radial velocity errors ($1\sigma$ deviation was used as the error of the period). The best fit sine curve is plotted on the radial velocity curve in Fig.~\ref{fig:rvs}.  The semi-amplitude of the optimal sine curve is $K = 155 \pm 6$ km/s. The top and bottom panel of Fig.~\ref{fig:ews} compares the behavior of the radial velocities and brightness of {\srgcv} using the photometric phases calculated from the Ephemerides (\ref{ephemeris}). The mean light curve in the bottom panel is the same in Fig.~\ref{fig:PDSbalman}, however unbinned in phase. The radial velocity curve is shifted relative to the sinusoidal light curve by $\Delta \varphi \approx 0.25$.

The behavior of the equivalent widths of several strong lines over the orbital period is shown in the Fig.~\ref{fig:ews} during the high state. The errors of EW were estimated by Monte-Carlo technique. The modulation of EWs with the orbital period is clear. All analyzed lines exhibit a single-peak equivalent width curve, except perhaps the He\,{\sc ii}$\sim$ $\lambda$4686 line, indicates a two-peaked structure. It should be noted that the behavior of EW of He\,{\sc ii}$\sim$ $\lambda$4686 line differs from hydrogen lines. Thus, the maximum of the EW of the hydrogen lines is near phase $\varphi \approx 0.35$, while the curve of the EW of ionized helium line reaches a maximum near $\varphi \approx 0.85$. It can be seen that the most dramatic EW variation ($\mathrm{[EW_{max} - EW_{min}] / EW_{min}}$) is with the He\,{\sc ii}$\sim$ $\lambda$4686 line about 60\%, while H$\beta$ and H$\gamma$ lines have EW amplitudes $\sim$45\%\ and $\sim$35\%, respectively. Fig. \ref{fig:ews} does not show any explicit correlation or anticorrelation between the EW of lines and the brightness of the system in photometric bands (i.e., in the continuum). Thus, the flux in the lines changes during the orbital period, which indicates a high optical thickness in the line emitting regions and/or obscuration by the donor (the latter requires high inclination $>$ 40$^{\circ}$). 

\subsection{The X-ray spectral and temporal data analysis}\label{sec:ana-xray}

\subsubsection{\nicer\ observation in the low state}\label{sec:ana-xray1}

In order to perform the \nicer\ analysis we used {\sc HEASoft}\footnote{https://heasarc.gsfc.nasa.gov/docs/software/heasoft/}\ (version 6.29) and utilized {\sc XSELECT}\footnote{https://heasarc.gsfc.nasa.gov/ftools/xselect/} (version v2.4-e) to obtain light curves and a spectrum using the cleaned events created via {\tt nicerl2} data reduction tool as described in Sec.~\ref{data:obs}. A background spectrum for our observation was created with the {\tt nibackgen3c50} tool and response and ancillary files were generated using the tasks {\tt nicerrmf} and  {\tt nicerarf}, respectively, to perform spectral analysis. Further analysis were conducted within {\sc HEASoft} environment  utilizing
{\sc XSPEC}\footnote{https://heasarc.gsfc.nasa.gov/docs/software/heasoft/xanadu/xspec/index.html}, for spectral and {\sc XRONOS}\footnote{https://heasarc.gsfc.nasa.gov/docs/software/heasoft/xanadu/xronos/xronos.html}, for temporal analyses. We underline here that the \nicer\ observation was  obtained in the low state after the source diminished by about 3 mag in the optical. During the construction and writing of the manuscript, another updated version of NICERDAS  was released in December 2022, NICER version-2022-12-16-V010a with calibration database xti20221001 including data product tasks {\tt nicerl3-spect} and {\tt nicerl3-lc} for spectral and light curve extractions. We have used the new software to check for consistency of results between different software versions and background checks. The results below will use the NICERDAS v.2020-04-23-V007a (and caldb xti20210707) unless an improvement have been found with the new version NICERDAS v.2022-12-16-V010a.

\begin{figure}
\begin{center}
\vspace{-0.3cm}
\includegraphics[width=8.8cm,height=5.0cm]{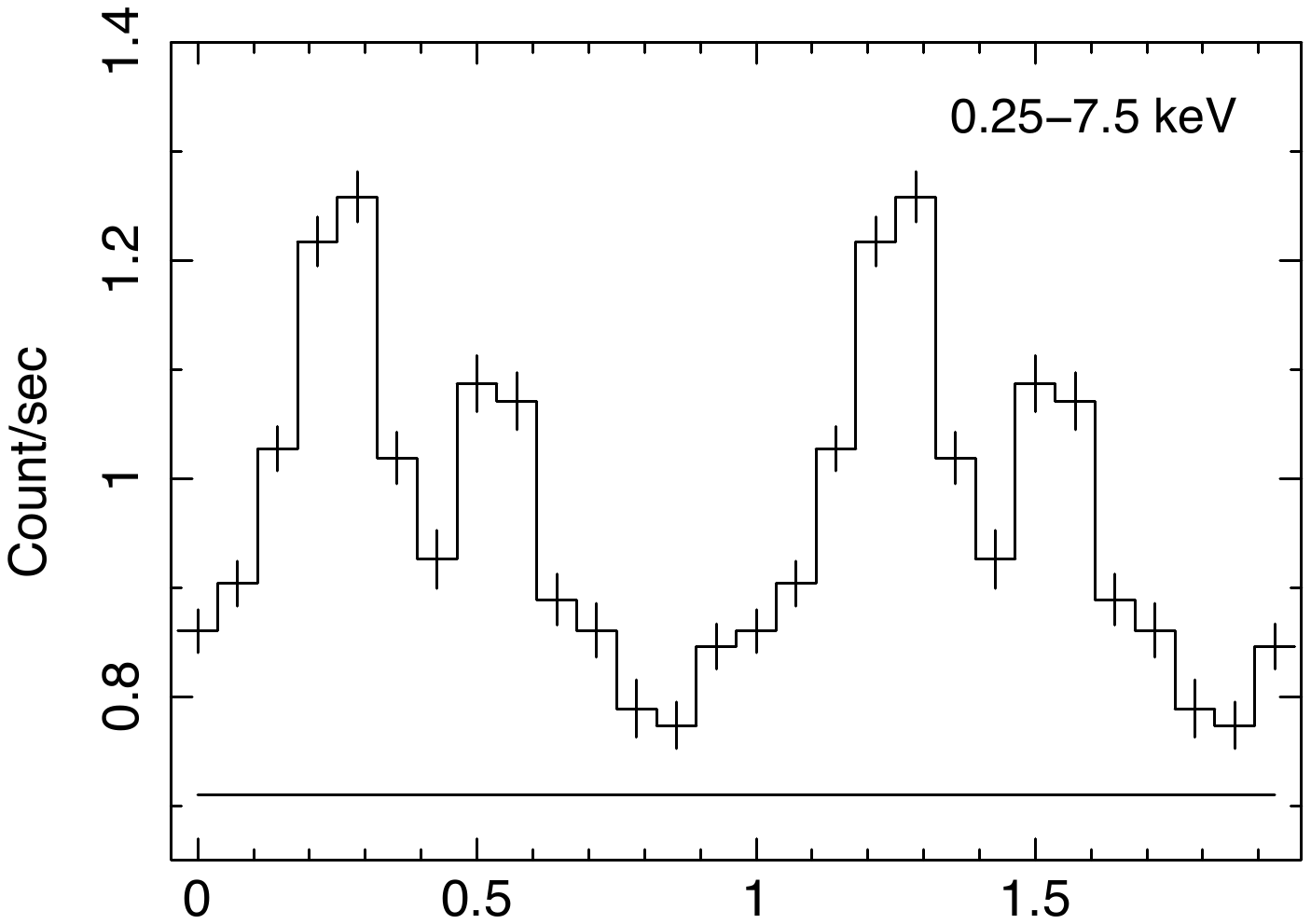} \\
\vspace{-1.25cm}
\includegraphics[width=8.8cm,height=5.0cm]{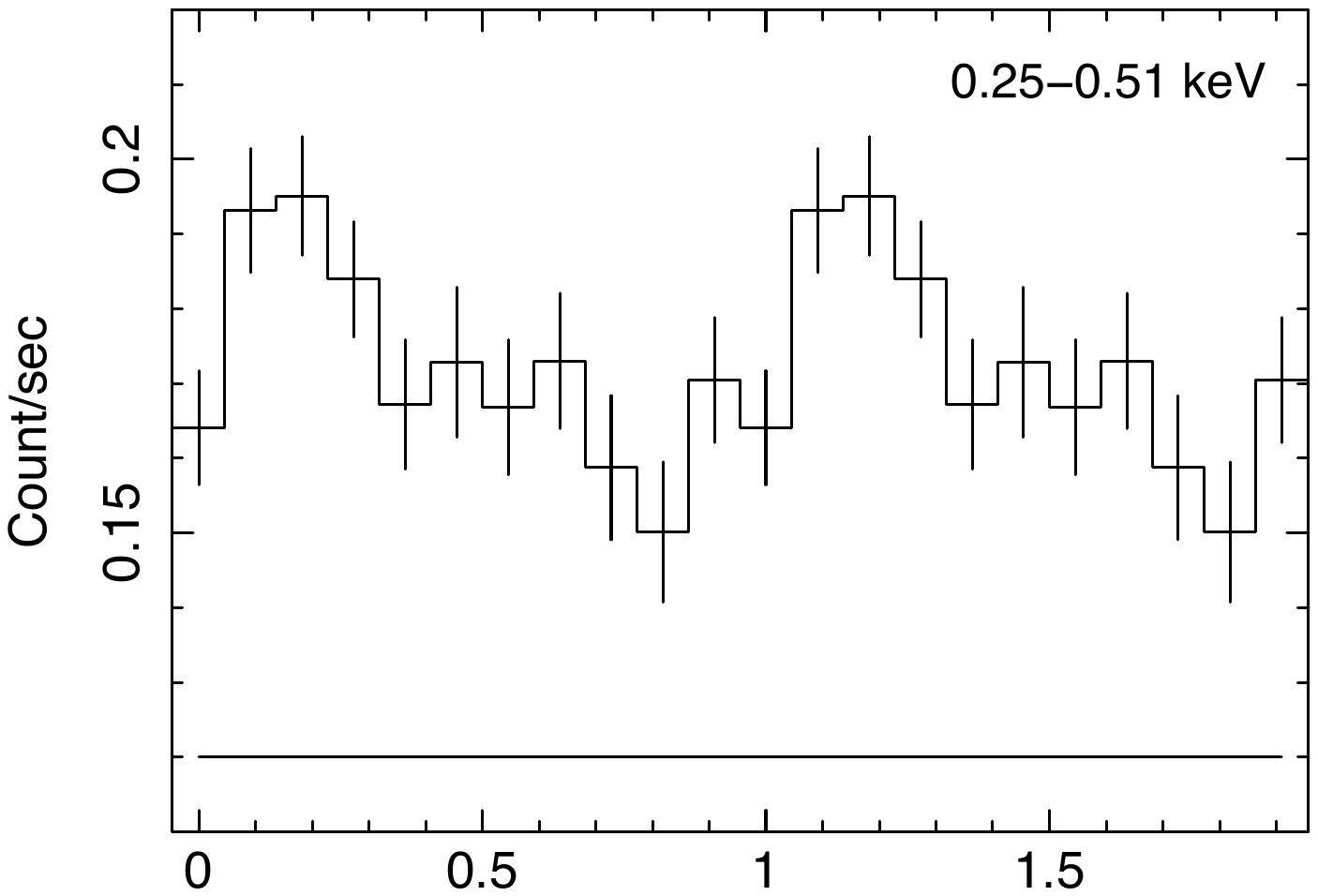} \\
\vspace{-1.25cm}
 \includegraphics[width=8.8cm,height=5.0cm]{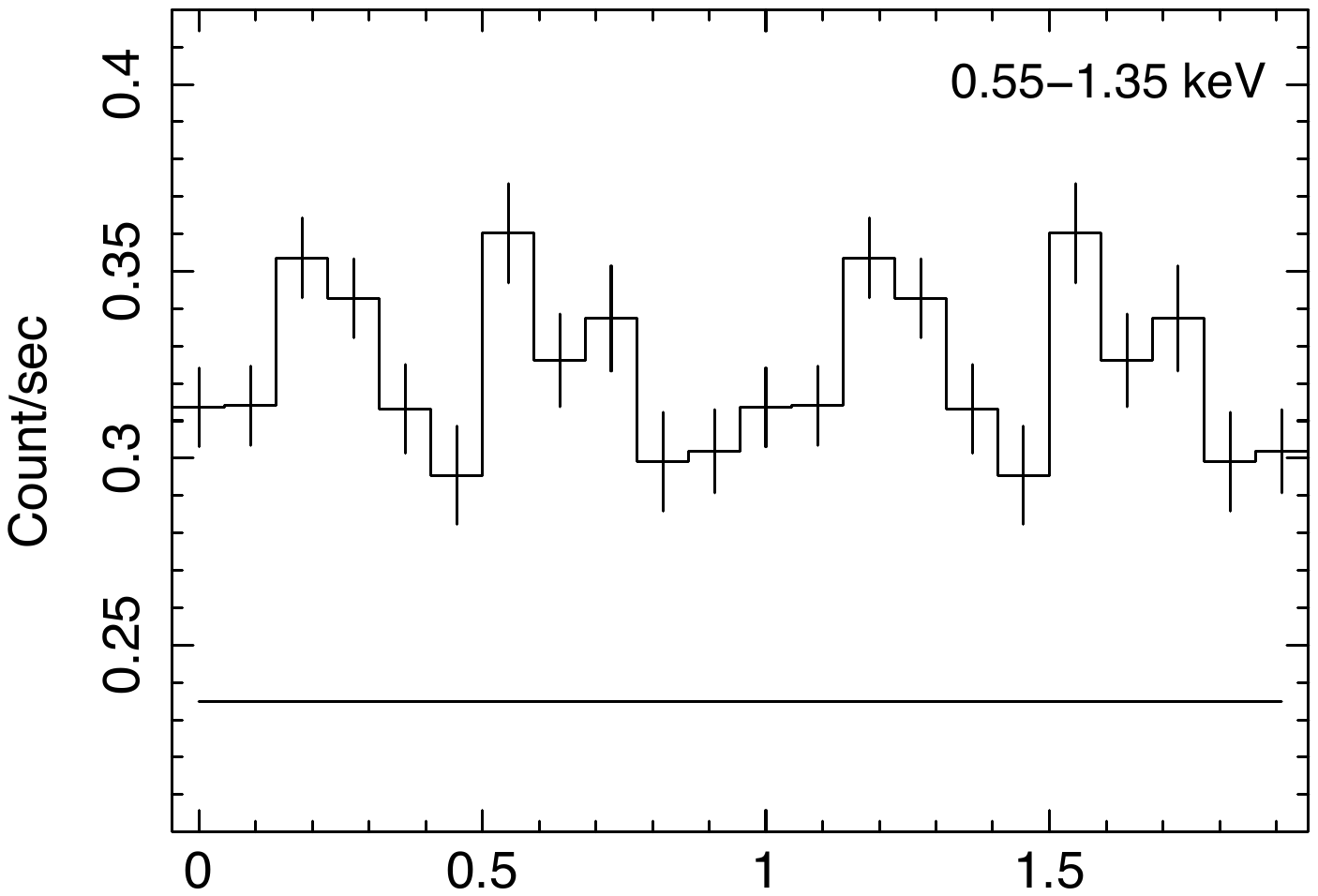} \\
\vspace{-1.25cm}
 \includegraphics[width=8.8cm,height=5.0cm]{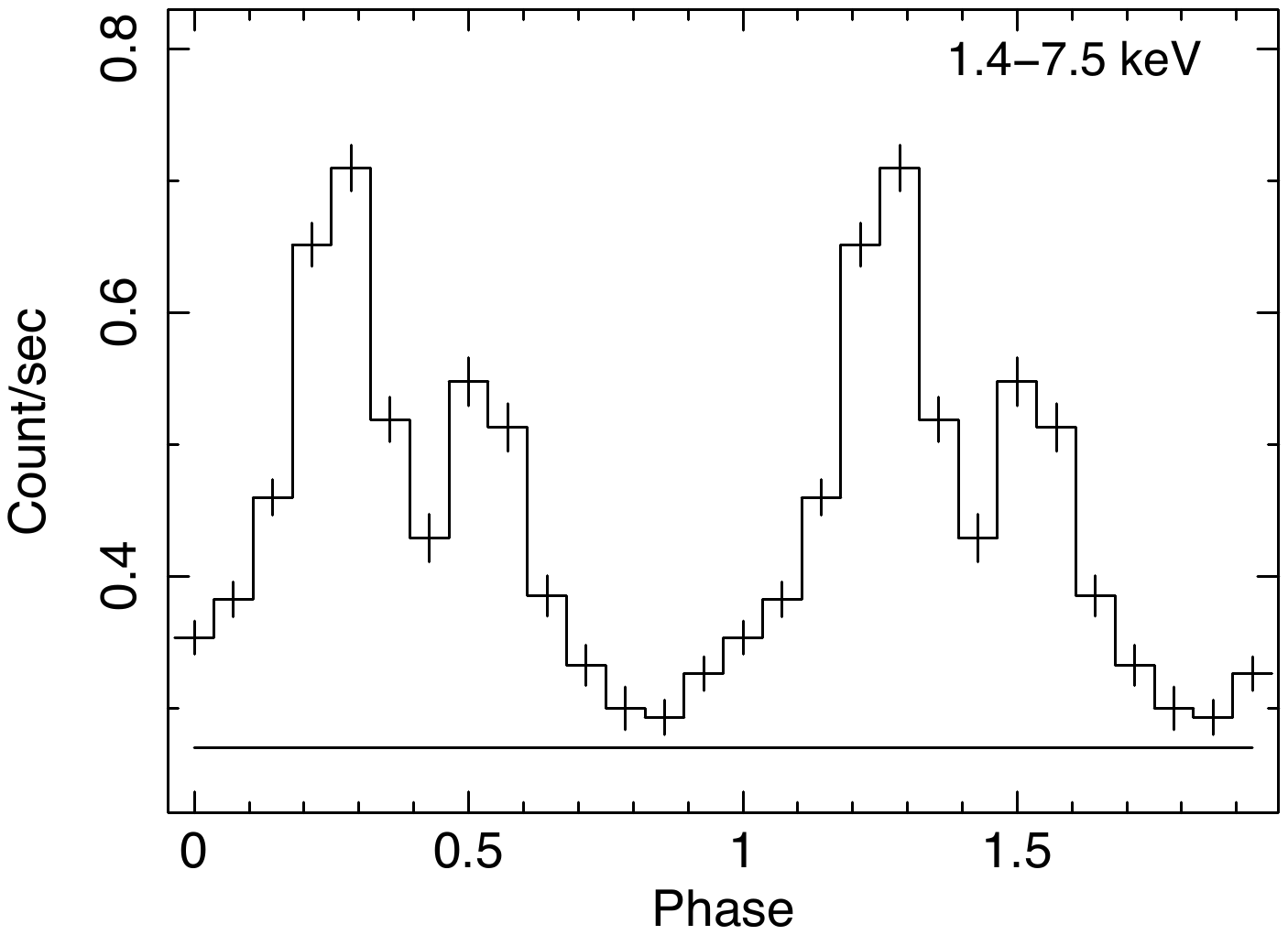} 
 
\vspace{-0.4cm}
    \caption{The \nicer\ pulse profile of the WD spin/orbital period in three different energy bands (\nicer\ 2021 June 13) calculated using light curves creatred with {\tt nicerl3-lc} task. The bands are labeled on the panels. The light curves are folded over the optical photometric Ephemerides in (1). The  background count rate calculated from the light curves for each band is denoted with the solid horizontal line.}
    \label{Balman:nicerlc}
   \end{center}
\end{figure}

Below we present the light curve analysis performed using  NICERDAS v.10a and {\tt nicerl3-lc} \footnote{ {\tt nicerl3-lc} task creates light curves from cleaned event files, calibrated and screened using standard criteria. The task produces background files over the given observation for inspection where we did not find any anomaly or high variability} task.
 After creating a total light curve in the \nicer\ range, a standard power spectral analysis was conducted which did not reveal any other periodicity. We folded the total X-ray light curve using the optical photometric Ephemerides we determined in this paper (the accumulated phase error at the time of the \nicer\ observation is $\sim$ 0.1). The resulting folded mean X-ray light curve between 0.25-7.5 keV is displayed in the top panel of Fig.~\ref{Balman:nicerlc}. This plot shows  variation of the X-ray flux at the orbital period (which for a polar-type MCV coincides with the WD spin period). 

We also computed pulse profiles in three  energy bands -- 0.25-0.5 keV, 0.55-1.35 keV, and
1.4-7.5 keV. They are shown in the three panels below the top panel in Fig.~\ref{Balman:nicerlc}.  The figure shows 
a  lack of strong modulation at the low energies below 1.4 keV and more significant
modulation in the 1.4-7.5 keV band, where the pulse profile has a double-peak structure indicating two-pole accretion in the system. An approximate estimate\footnote{There is mostly a constant background in the folded light curves (denoted according to the energy band) in  Fig.~\ref{Balman:nicerlc} ($\sim$ 1.0 c s$^{-1}$ in 0.2-10.0 keV) which resembles to rates of the blank sky fields of the Rossi X-ray Timing Explorer Observatory.} of pulsed fraction, (F$_{max}$--F$_{min}$)/(F$_{min}$+F$_{max}$), in this energy range is  $\sim$ (30-40)\%. The estimates for the other ranges are low $\sim$ (10-15)\% for 0.25-0.5 keV and $\sim$ 10\% for the 0.55-1.35 keV range. 
The pulse profiles (mostly sinusoidal) in the optical and X-ray bands do not show evidence for an eclipse.
For example, a two-pole accreting polar detected in the X-rays, V496 UMa, shows similar double-peaked  pulse profiles to ours \citep{2022Kennedy,2022Ok}. 
We have also clarified that the 92.6 min orbital period of the spacecraft does not affect the modulation detected at the WD spin period or affect the light curve modulations in different energy bands.

\begin{figure}
\vspace{-0.5cm}

\hspace{-0.2cm}
\vspace{-0.85cm}
\includegraphics[width=8.95cm,height=6.3cm]{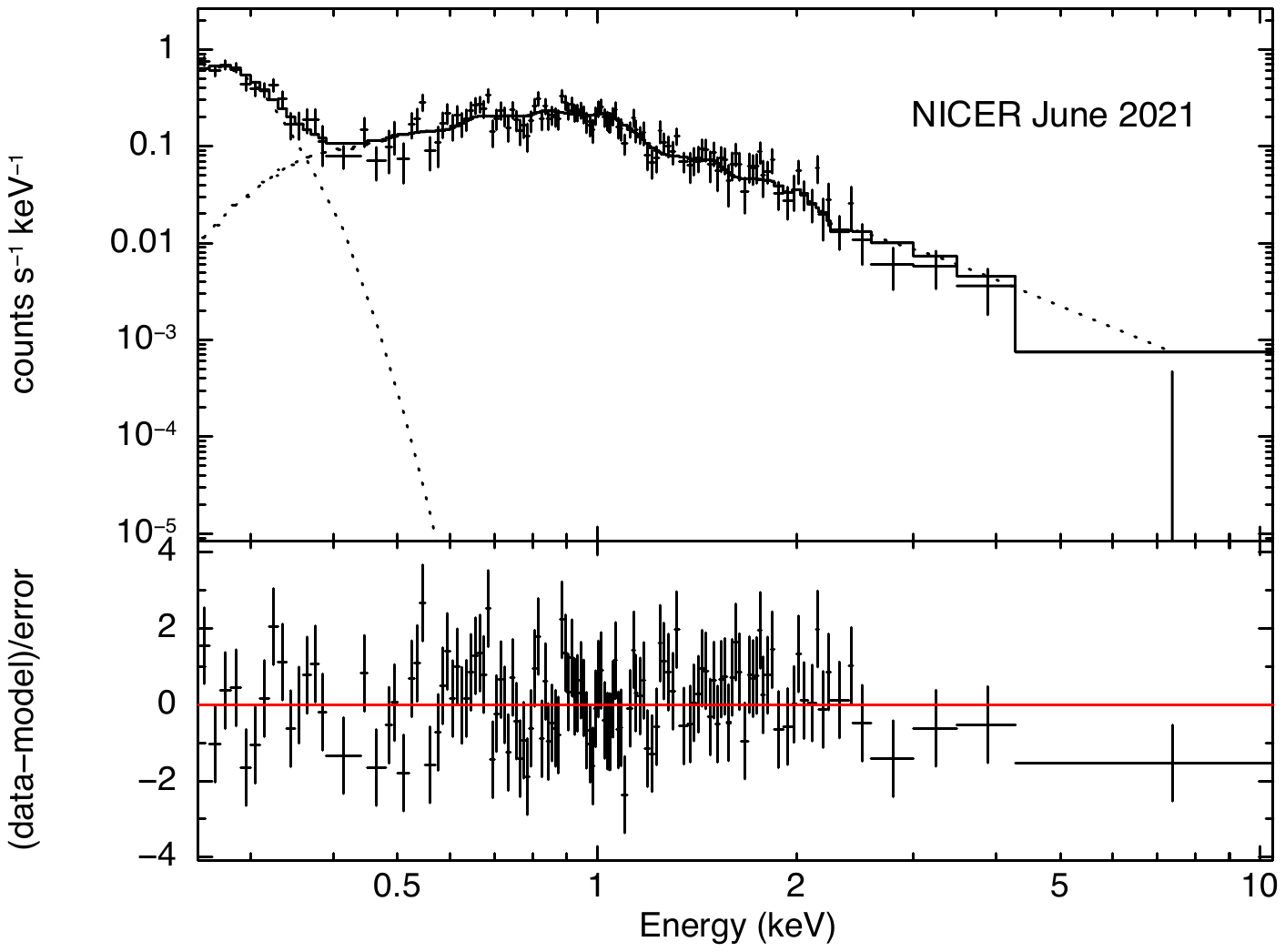} 
\vspace{-0.85cm} 
\hspace{0.17cm}
 \includegraphics[width=8.8cm,height=5.3cm]{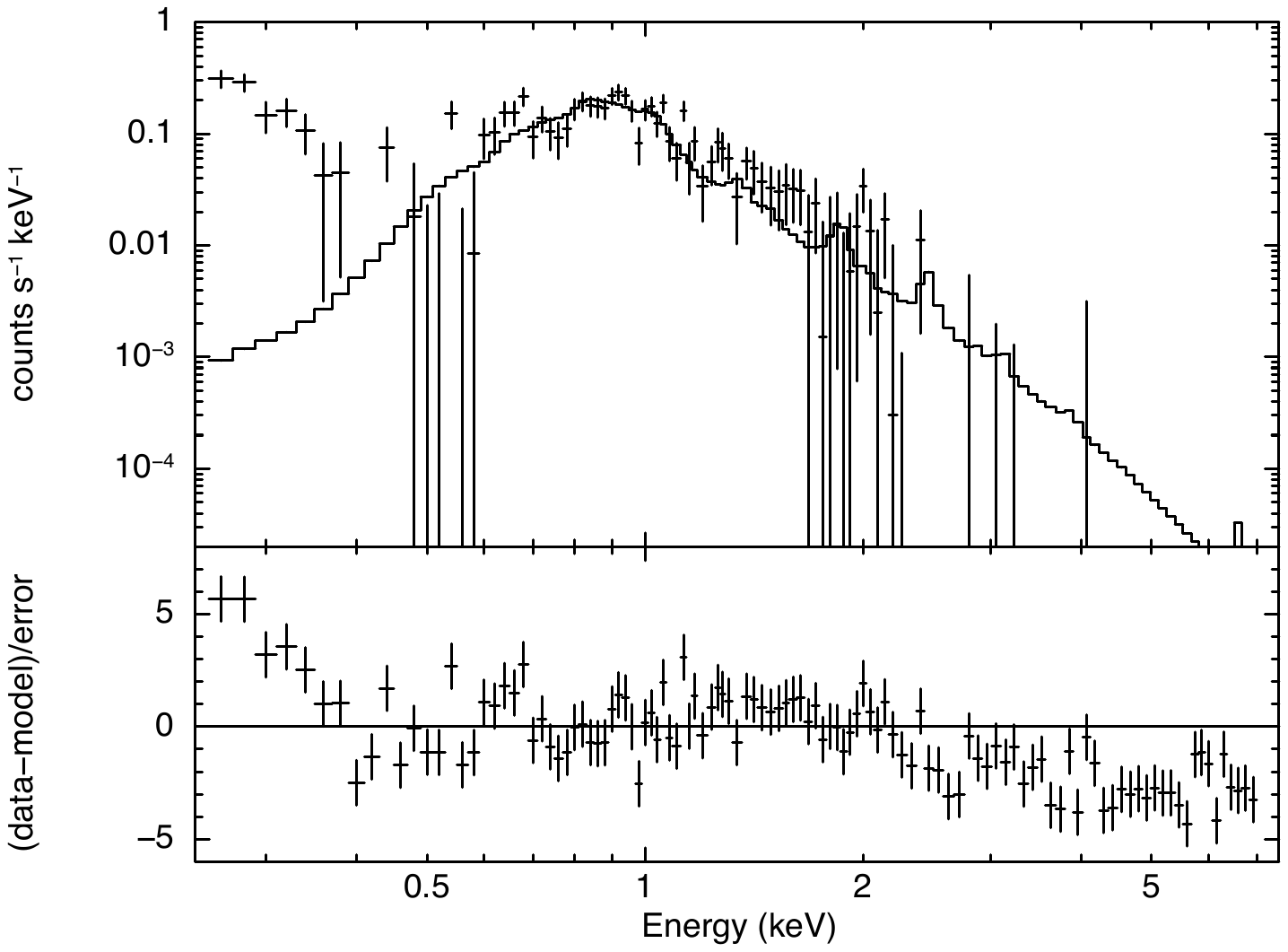} 
 \vspace{0.05cm} 
 \hspace{0.2cm} 
 \includegraphics[width=8.8cm,height=5.4cm]{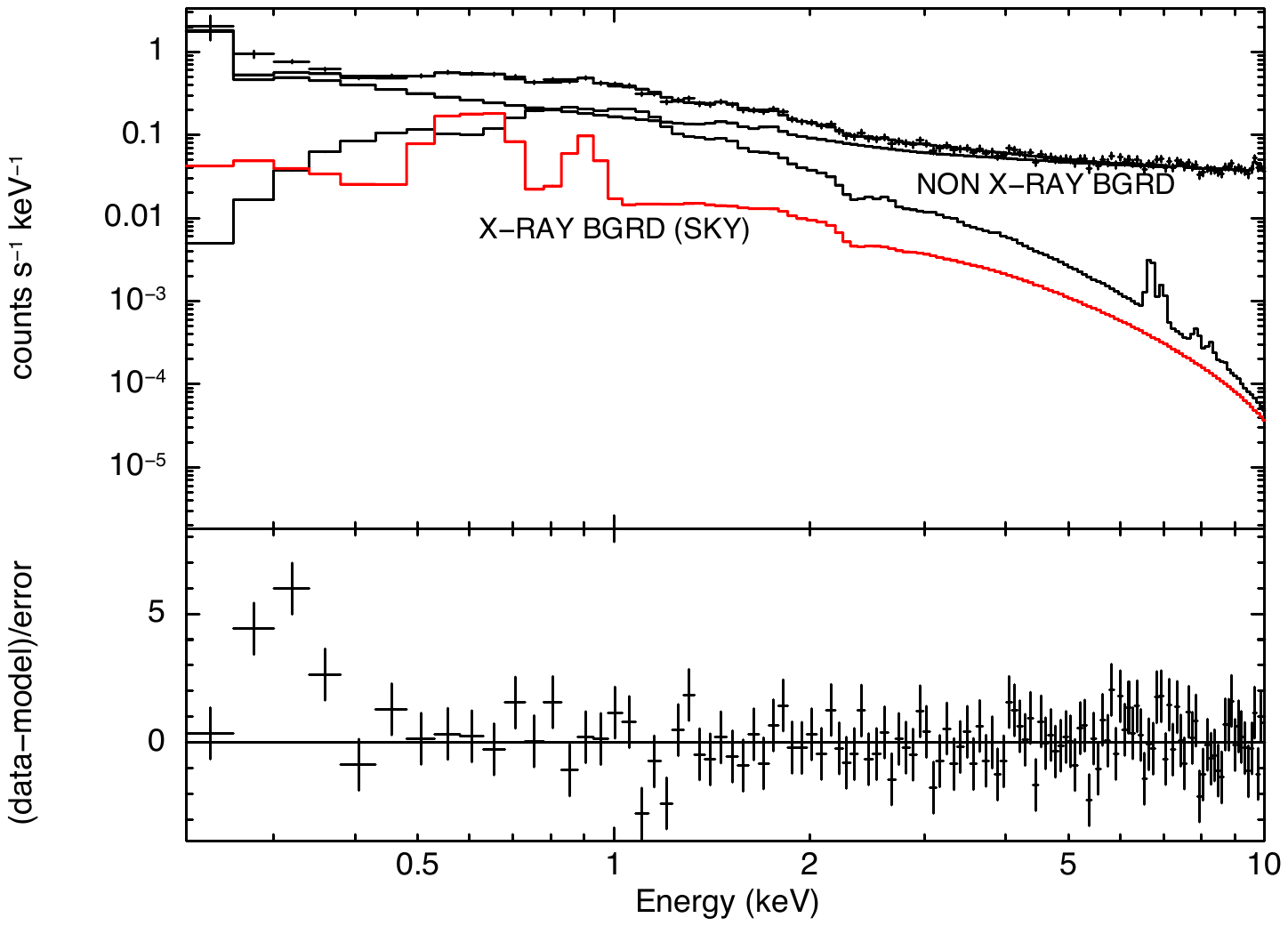} 
 
\vspace{-0.3cm}
    \caption{
    Top panel shows the X-ray spectrum of \srgcv\ obtained with \nicer\ between 0.25-10.0 keV (using 3C50 background model). The spectrum is fitted with {\sc tbabs$\times$(bbody+cevmkl)} labeled with dotted lines and the residuals in the lower part are in standard deviations for all panels. The middle panel is a spectrum derived using a very stringent undershoot parameter range 0-50 to clear the dominant soft X-ray noise peak below 0.5 keV, fitted with only a CEVMKL model using the 3C50 background model. The bottom panel shows the same spectrum created using the SCORPEON background model with different variable background contributions, fitted without the blackbody component (see text for details).}
    \label{Balman:nicersp}
\end{figure}

After creating the spectrum (using {\sc XSELECT}), and calculating a background spectrum via the tool {\tt nibackgen3c50}, these have been incorporated in {\sc XSPEC} along with the proper ancillary and response files for spectral modeling and statistical analysis.
The background estimator, {\it nibackgen3c50}, models both the non-X-ray and the X-ray sky background contributions.
There is a clear soft excess in the NICER spectrum (Fig.\ref{Balman:nicersp}) and  we used {\sc tbabs$\times$(bbody+cevmkl)} composite model for deriving the spectral parameters that describe the source emission in the 0.25-10.0 keV range. BBODY is a blackbody model of emission and CEVMKL \citep{1996Singh} is a model of multi-temperature plasma emission in collisional equilibrium with variable plasma abundances build from the MEKAL code \citep{1986Mewe}. The fits are performed using interpolation on pre-calculated MEKAL tables with abundances fixed at solar values. However, fits and error ranges are checked against the same code running using the ATOMDB\footnote{http://www.atomdb.org}~ database.  Top panel of Fig.\ref{Balman:nicersp} displays the fitted \nicer\ spectrum using the 3C50 background model. This is in accordance with the expectations of emission from the stand-off shock near the WD surface in the accretion column. Note here that there is no disk and the accreting material is carried along the magnetic field lines via an accretion funnel.  

The two-component fits to the \nicer\ spectrum (in top panel Fig.\ref{Balman:nicersp}) yield an \nh\ of (0.5-0.95)$\times$10$^{21}$ cm$^{-2}$ using abundances set to "wilm" \citep{2000Wilms} (90\%\ confidence level range for a single parameter). The Galactic absorption  in the direction of the source is (6.46-6.74)$\times$10$^{21}$ cm$^{-2}$ calculated via the {\tt nhtot}\footnote{https://www.swift.ac.uk/analysis/nhtot} tool devised using the GRB data of \swi\ Observatory \citep{2013Willingale}. The  \nh\ value from the spectral fit is about 5-10 times lower than the interstellar absorption in the direction to the source. This suggest that the source is located in front of the main absorbing material in this direction (e.g., source is closer by) and that the intrinsic absorption is also, low with perhaps some partial covering and ionized absorption in effect.

The other spectral parameters of our modeling yield parameter ranges  using the CEVMKL model with the power spectral index of the temperature distribution {$\alpha$}=1 are : 1) kT$_{max}$=3.0-6.0 keV, 2) N=(4.0-6.1)$\times$10$^{-4}$, 3)   kT$_{\rm BB}$=0.015-0.018 keV, 4) N$_{\rm BB}$=(0.3-2.4)$\times$10$^{-2}$.  When ${\alpha}$ is not fixed (for CEVMKL), these spectral parameters change to 1)    kT$_{max}$=7.4-13.5 keV, 2) N=1.7-4.7$\times$ 10$^{-4}$, 3) kT$_{\rm BB}$=0.015-0.019 keV, 4) N$_{\rm BB}$=(0.6-4.5)$\times$10$^{-2}$ with an ${\alpha}$ in a range 0.5-0.7. All error ranges correspond to 90\%\ confidence level for a single parameter.
N$_{\rm BB}$ is the normalization of the BBODY and N is the normalization of CEVMKL model where N=(10$^{-14}$/4$\pi$D$^2$)$\times$EM and EM (emission measure) =${\rm \int n_e\ n_H\ dV}$ (integration is over the emitting volume V). Note that the spectral data products and these parameter ranges are consistent for different NICERDAS software versions (v.7 and v.10a) as the fits were checked using both.

The absorbed flux derived from the fits is about 5.0$\times$10$^{-13}$\fluxcgs\ in the 0.2-10.0 keV band. The soft X-ray and the hot plasma components have absorbed fluxes of 1.0$\times$10$^{-13}$\fluxcgs\ and 4.0$\times$10$^{-13}$\fluxcgs, respectively in the 0.2-10.0 keV band. The unabsorbed fluxes, in the same energy band, at the 90\% confidence level are (6.0$\times$10$^{-13}$--2.0$\times$10$^{-12}$) \fluxcgs\ (for the soft component) and  (1.7$\times$10$^{-13}$--5.8$\times$10$^{-13}$) \fluxcgs (for the hot plasma), respectively.

The ${\alpha}$ parameter of 1.0 signifies a standard cooling flow plasma in collisional equilibrium. The change of this value from 1.0 indicates that the flow deviates from a cooling flow and that the flow does not cool effectively. The temperature range derived from the variable ${\alpha}$ value is higher with plasma temperatures 7.4-13.5 keV, as expected.
We find that the fit with no fixed ${\alpha}$ parameter ($\chi^2$=279.9 dof=249) is better than the fit with ${\alpha}$=1.0 ($\chi^2$=290.0 dof=250) at 99.99\%\ Confidence Level, over 3$\sigma$ significance (using an FTEST in {\sc XSPEC}). If ATOMDB (database) is used for calculating the plasma model, the discrepancy in total $\chi^2$ yields a 90\%\ confidence level difference between the fits.

The \nicer spectrum of SRGA J213151.5+491400 shows prominent soft component in its spectrum  below $\sim 0.5$ keV (0.25-0.5 keV). To our knowledge, this the first time such a component is detected in the low state of a polar. Below we discuss and eliminate several contaminating factors which may affect the reliability of its detection.

\nicer\ spectra are contaminated by the low energy noise peak below 0.25 keV resulting from undershoots\footnote{https://heasarc.gsfc.nasa.gov/docs/nicer/analysis$\_$threads/undershoot-intro/} (strongest background component). We have checked how much this may affect our detected soft X-ray component. To this end  we used  a highly conservative constraint on undershoot count range ($\sim$ 0-50) in data screening that minimizes the low energy noise contributions to almost zero. This method yielded a (5-7)$\sigma$ excess in the residuals between 0.25-0.5 keV while
the neural H column density parameter was set free (see middle panel of Fig.\ref{Balman:nicersp}). The \nicer\ calibration team (\nicer\ SOC), cordially, checked any anomaly in the background and noise characteristics of our data and the observation confirming the soft X-ray component of the source. 

In order to check the possible contribution of soft X-ray background to the soft component found in the \nicer\ spectrum of SRGA J213151.5+491400, we also used the SCORPEON background model  (using  NICERDAS v.2022-12-16-V010a. and {nicerl3-spect} task) where one can obtain composite (modelable) background-components (produced via a script) along with the data products to use in the fitting process. Our fits revealed that utilizing the SCORPEON model, the spectral parameter ranges for our source are consistent with the results using varying $\alpha$ parameter  and slightly higher temperature in the above paragraph (an $\alpha$ of 1 is acceptable within the parameter space) . Firstly, we checked the soft X-ray contribution from the Local Hot Bubble (LHB) and varied the emission measure parameter of LHB out to 50-100\% of the best fit value.  The best fit value of 1.6$\times$10$^{-3}$\ cm$^{-3}$ pc is consistent with the sky maps presented in \citet{2017Liu} around the coordinates of our source (solar charge exchange contribution subtracted) and the enhancement we used is larger than permitted by values on these maps. We found that LHB can not account for the soft excess below 0.5 keV yielding (5-6)$\sigma$ variation in the fit residuals (see
bottom panel Fig.\ref{Balman:nicersp}, the red line for total X-ray sky background).  We also varied the SWCX (solar wind charge exchange) sky background parameters (the charge exchange line emission by the heliosphere) to account for the expected  lines (e.g., O K$\alpha$, O {\sc vii}, O {\sc viii}, Ne {\sc ix} -- 0.5-1.0 keV) in the background emission which also reduced residual fluctuations (see bottom panel of Fig.\ref{Balman:nicersp}, the red line between particularly 0.5-0.9 keV) . We derived best-fit line fluxes as, 2.2 (O K$\alpha$), 5.4 (O {\sc vii}), 6.4 (O {\sc viii}), and 2.7 (Ne {\sc ix}) in LU (line units, photons s$^{-1}$cm$^{-2}$ sr$^{-1}$). 
During these fits the LHB background was fixed at the 50\% larger value of the best fit. The SWCX line fluxes found from the fits are large/enhanced compared to the average maximum permitted values of 4.6 LU for O {\sc vii} and 2.1 LU  for O {\sc viii} (see \citet{2007Kout}, our values are 1.2 times larger for O{\sc vii} and 3 times larger for O {\sc viii}). We note that the maps in \citet{2017Liu} and surveys therein including the \rosat\ RASS (all sky survey), indicate that the field around our source is empty with no sign of excessive diffuse emission in the vicinity. Therefore,  bottom panel of Fig.\ref{Balman:nicersp} shows that  even when highly elevated values of LHB and SWCX sky-backgrounds are assumed, there is a soft X-ray excess at (5-6)$\sigma$ as observed in the residuals that can not be modeled by background effects. 

We note here that the field of view of \nicer\ is 5\pri. 
We have checked the X-ray position of our source for unrelated sources using all available archival X-ray data, X-ray surveys and X-ray catalogues. We found no other X-ray source in the field of view of \nicer\ (this is why it was proposed). This is larger
than the error box we have used for optical identification. We stress that in the 1\pri\ search diameter, where we found 30 optical counterpart candidates down to 20-21 magnitudes (and derived their spectra), we had no other source that 
was blue enough and showed as an interacting system that would be emitting X-rays than this one source \gcv.  Particularly, there was no indication of a strong UV candidate that would yield the $\sim$ 15 eV blackbody detected by \nicer\ other than \gcv.
Moreover, neither \erosita\ (HPD$\sim$ 15\pri\pri) and nor \artxc\ (HPD$\sim$ 30\pri\pri) imaging surveys detected any other X-ray source in the vicinity of the new source within the 5\pri\ FOV of  \nicer\ during the four scan periods in 2020 and 2021. We have checked all \gaia\ alerts and there were no other transients in the 5\pri vicinity of \gcv\ detected after its raise in the years 2020 and 2021.

\subsubsection{\erosita\ observations in the high and low states}\label{sec:ana-xray2}

\srgerosita\  detected \srgcv\ in each of the four all-sky surveys performed by the \spektr\ observatory in 2020-2021. \erosita\ observations took place during the following time intervals: 2020 June 2-4, 2020 December 6-8, 2021 June 6-8, and 2021 December 10-12.  \erosita\  provides good quality spectral information in the 2020 high state of the source not covered by \nicer\ observations. 

The \erosita\ raw data were processed by the calibration pipeline developed in the \erosita\ X-ray catalog science working group at Space Research Institute (IKI, Moscow, Russia) using the calibration tasks of the  \erosita\  Science Analysis Software System ({eSASS}) developed at Max-Planck Institute for Extraterrestrial Physics (Garching, Germany) and the data analysis  software developed in the RU eROSITA consortium at IKI. We excluded from the analysis telescope modules 5 and 7 affected by the optical light leakage. The sources spectra and light curves were extracted using the circular aperture with radius of $60\arcsec$ centered at the source position. An annulus region with the inner and outer radii of $120\arcsec$ and $300\arcsec$ was used for the background estimations. The response files were created using the {\tt eSASS} tasks. Spectral and temporal analysis were performed using {\sc XSPEC} and {\sc XRONOS} packages within the same {\sc HEASoft} environment as in \ref{sec:ana-xray1}.

For the high state, we fitted simultaneously the two  spectra collected during the 1-st and 2-nd all-sky surveys on 2-3 June and 6-8 Dec 2020. All model parameters were linked for the two spectra, except for normalizations. A single component  {\sc XSPEC} model {\sc tbabs$\times$cevmkl} provided adequate description of the observed spectra in the high state given the spectra statistics.  
Similar to the approach used in the \nicer\ data analysis, we performed two types of fits fixing the $\alpha$ parameter of CEVMKL model at 1 or making it a free parameter of the fit. We used the CEVMKL model with the ATOMDB database.  For {$\alpha$}=1, we obtained the following spectral parameter values: 1) kT$_{\rm max}$=76.8$_{-51.8}^{{+}\infty}$ keV, 2) N=(8.0$_{-2.1}^{+4.6}$)$\times$10$^{-3}$ and N=(5.4$_{-1.4}^{+3.2}$)$\times$10$^{-3}$ for the 1st and 2nd survey respectively. N is the normalization of the CEVMKL model (see Sec.~\ref{sec:ana-xray1} for the description). The \nh\ parameter
was $0.6_{-0.3}^{+0.6}\times$10$^{21}$ cm$^{-2}$ consistent with the \nicer\ value.  Similar to \nicer\ spectral fits, this value is almost an order of magnitude smaller than the Galactic value.  The C-Statistical value of the fit is 109.6 for 110 degrees of freedom (using Survey 1 and 2 spectra). The plasma temperature in the accretion column is unconstrained with the lower limit of kT$_{max} >$ 21 keV at the 95\% confidence  level (using ATOMDB). The lower limit changes to  kT$_{max} >$ 17 keV when using an interpolation on pre-calculated MEKAL tables. (see Sec.~\ref{sec:ana-xray1}). For ${\alpha}$  thawed, we obtained only an insignificant reduction of the C-Statistics (= 102.7 for 109 degrees of freedom) and obtained a best fit value of $\alpha$=1.3 with  a lower limit of $\alpha>0.7$ (95\% confidence  level). 

\begin{figure}
\vspace{-0.5cm}
	\includegraphics[width=9.2cm,height=7cm]{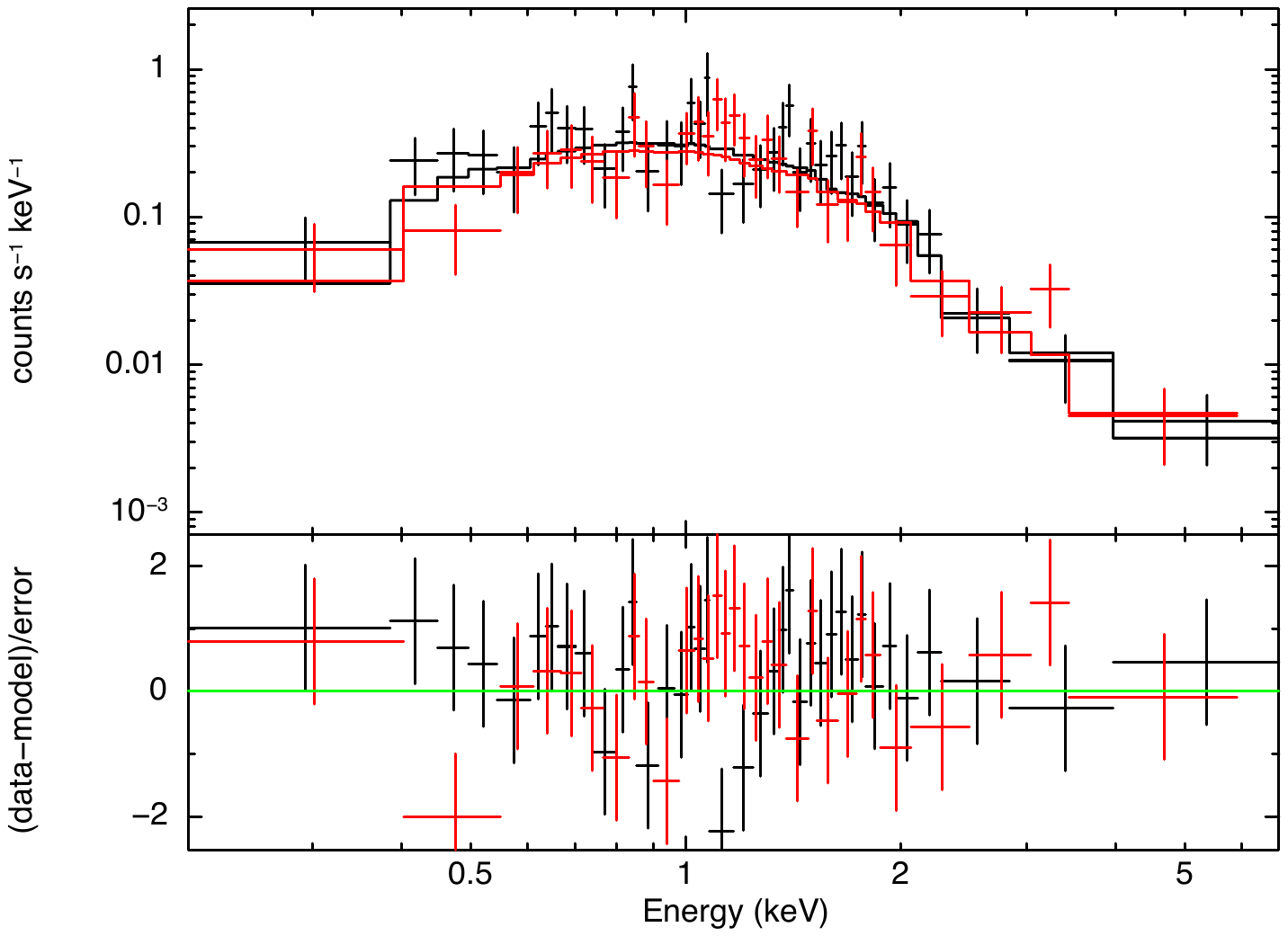}
    \caption{The X-ray spectra of \srgcv\ obtained with \srgerosita\ telescope in the course of two sky surveys in 2020 (high state). The lower  panel shows the residuals in standard deviations.  
    }
    \label{marat:erositasp}
\end{figure}

Given no improvement in the fit statistics, and an $\alpha$ of 1.0 is acceptable within the error range, a standard cooling flow model can be used for the physical interpretation of the high state. Fig.~\ref{marat:erositasp} displays the  spectra  obtained in the 1st and 2nd surveys along with their best-fit model (using $\alpha$=1). As one can see, in the high state there is no soft component as apparently present as in the  low state spectrum obtained with \nicer\ (cf. Fig.~\ref{Balman:nicersp}). However, the CEVMKL continuum level is about an order of magnitude higher in the high state, and the presence of a soft component with the parameters measured  by \nicer\ is consistent with the \erosita\ data.  The 90\% upper limit on the absorption corrected flux (0.2-10.0 keV) of a blackbody component with the temperature equal to a \nicer\  best-fit value of 15 eV is  7.6$\times$10$^{-11}$\fluxcgs\ larger than  the unabsorbed flux range measured by \nicer\ in the same energy band. 

The two spectra obtained by \erosita\ in the 3rd and 4th sky-surveys in 2021 have relatively low number of counts (about $\sim 25-30$ counts each) and do not provide statistically significant spectra in the low state, being generally consistent with \nicer\ results. The absorbed fluxes of the best fitted models measured in the four \srgerosita\ surveys are: 
$(6.3\pm0.8)\times$10$^{-12}$ \fluxcgs, $(4.3\pm0.6)\times$10$^{-12}$ \fluxcgs, $6.9_{-2.4}^{+3.0}\times$10$^{-13}$ \fluxcgs\ and $6.9_{-2.2}^{+2.6}\times$10$^{-13}$ \fluxcgs\ in the 0.2-9.0 keV band. 


\section{Discussion}

\subsection{Doppler Tomography in the high state}

The behavior of emission line profiles as a function of the  orbital phase is analyzed by the means of Doppler tomography technique which aims to reconstruct the distribution of emission regions in the two-dimensional velocity space. Each point in this space has two polar coordinates: $v$ and $\vartheta$. $v$ is the absolute value of the velocity (relative to center of mass) projected to the line of sight and $\vartheta$ is the angle between the velocity vector and the direction to the secondary as it is seen from the center of mass (see \cite{kotze15} and \cite{2016Kotze} for details). In this work Doppler tomograms were reconstructed by doptomog code \citep{kotze15}, which implements maximum entropy regularization technique. Moreover, we used the so-called flux-modulated variant of Doppler tomography, which assumes sine-like variability of emission points during orbital period. This option is preferable for studying an optically thick media \citep{steeghs03}. The Doppler tomograms can be reconstructed in two projections: standard and inside-out. In the standard projection, the absolute value of velocity $v$ increases from the origin of the tomogram to the periphery which is a good choice for analyzing low-velocity structures, the high-velocity structures are blurred. The inside-out projection eliminates this problem using an absolute velocity that increases from periphery to the center of the tomogram \citep{kotze15}. 
Based on the available observations, it is impossible to construct orbital ephemeris for \srgcv. The system is not eclipsing, and the spectra do not show lines (or line components) formed on stellar components. Due to the absence of orbital ephemeris, we have assumed that orbital phases $\varphi_{orb}$ (where $\varphi_{orb} = 0$\  is taken as the closest approach of the secondary to the observer) are related with photometric phases $\varphi$ as $\varphi_{orb}$ = $\varphi$ $+$ 0.76\ . Using this relation, we achieved the best agreement between the emission distribution on the Doppler tomogram and the theoretical accretion stream velocities.  The regularization parameter used in maximum entropy method was selected by L-curve criterion (see, for example, \cite{hansen93}). 

\begin{figure*}
\centering{
	\includegraphics[width=0.84\textwidth]{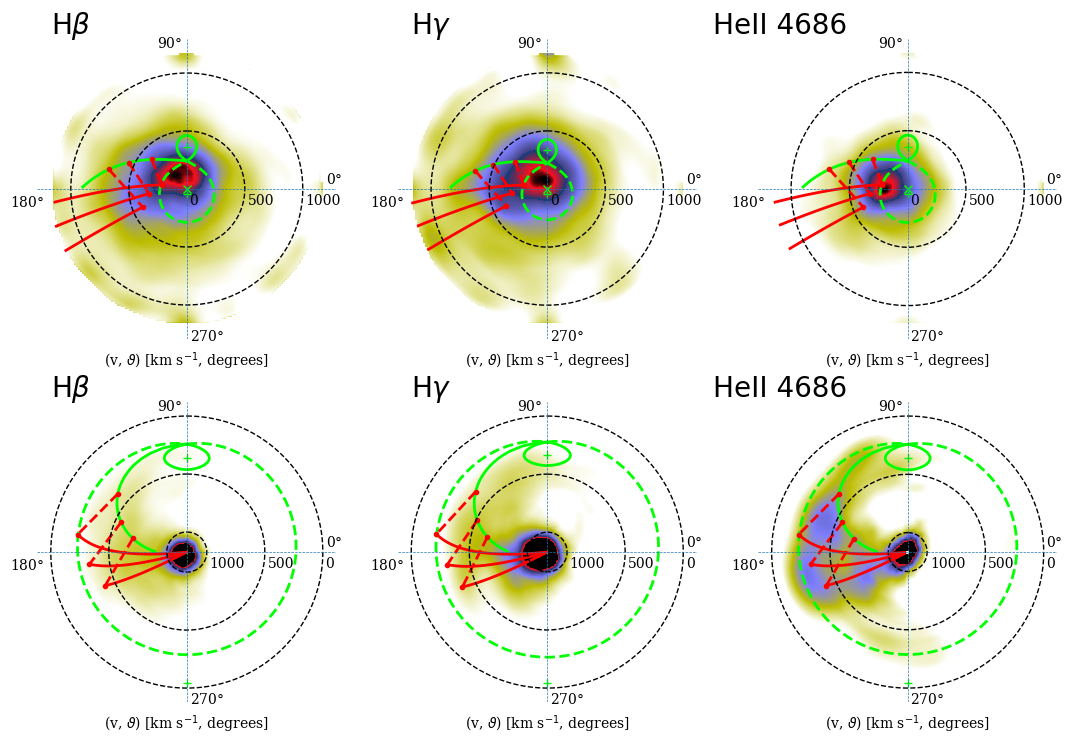}
    \caption{ Upper panel: Doppler tomograms for H$\beta$, H$\gamma$ and He\,{\sc ii}~$\lambda$4686 in standard projection. Lower panel: the same as in upper panel but in inside-out projection. Closed dashed line and closed continuous lines are velocities of primary and secondary Roche lobes, respectively. The {\bf green} line is the trace of ballistic stream in velocity space. The {\bf red} lines are velocities of a gas stream at three dipole lines.}\label{fig:dopptom}}
\end{figure*}

The Doppler tomograms were reconstructed using the spectral data obtained with VPHG1200G grism on BTA/SCORPIO. The tomograms in standard projection for H$\beta$, H$\gamma$, and He\,{\sc ii}~$\lambda$4686 lines are presented in the upper panel of Fig.~\ref{fig:dopptom}. They are one-spotted and the position of spots is clearly different for He\,{\sc ii}~$\lambda$4686 and Balmer lines. Also, there is slight elongation of emission spots which may point out to the direction of accretion flow. Any ring-like structure that could correspond to an accretion disc doesn’t appear in tomograms which confirms the polar nature of \srgcv. 
The Doppler tomograms using inside-out projection are presented in lower panel of Fig.~\ref{fig:dopptom}. The tomograms  show  the high-velocity ($v>500$~km/s) component of the accretion stream directed to the center. The possible separation of the stream into two components is noticeable in the tomogram for He\,{\sc ii}~$\lambda$4686 at $500\lesssim v \lesssim 750$~km/s and $135 \lesssim \vartheta \lesssim 200^{\circ}$. This separation can be caused by the transitions of the accreted gas from the ballistic to the magnetic trajectory. For a demonstration of this effect, we superimposed on the tomogram the theoretical velocities of particles starting from the Lagrangian point L$_1$. At first, the particles move along a ballistic trajectory calculated by solving the restricted three-body problem. Then they abruptly switch to a magnetic trajectory described by a dipole model. To perform these calculations, we set the secondary mass $M_2 = 0.1 M_{\odot}$ satisfying the empirical relation $M_2/M_{\odot} = 3.453 P_{orb}^{5/4}$ of \cite{1995Warner}. The mass of the white dwarf is taken to be $M_1 = 0.83 M_{\odot}$ as the most probable primary mass in cataclysmic variables \citep{2020Zorotovic}. The inclination is fixed at $i=50^{\circ}$ by fitting the theoretical ballistic trajectory to observed one on the Doppler tomogram. The magnetic trajectories are calculated for three azimuth angles ($10^{\circ}, 20^{\circ}, 30^{\circ}$) which determines the position of ballistic-to-magnetic trajectory transition region. Signs of stream separation into two components (ballistic and magnetic) in Doppler tomograms are typical for polars \citep{schwope97}. Other signs of stream separation into two components recovered from inside-out projections are V834~Cen \citep{2016Kotze}, BS~Tri \citep{2022Kolbin}, V884 Her \citep{1999Hastings, Tovmassianetal2017}.

\subsection{Comparison of X-ray emission in different source states}

Polars (or AM Her systems) unlike non-magnetic CVs or intermediate polar systems, do not have a disk and a reservoir of material for that purpose. As a result when mass transfer ceases, accretion stops.  In  high accretion states (${\dot{m}}\ge 1\  g~s^{-1} cm^{-2}$) the density of the accreting gas in the accretion column is high to maintain a collisional time-scale shorter than the cooling and dynamical time-scales, hence the flow is hydrodynamic and a stand-off shock forms in the accretion column near the WD surface forming the hard X-ray component and the soft X-ray component forms via the reprocessing of the hard component \citep{1975Basko,1979Lamb}.  A widely accepted model for the soft X-ray component during the high state is the "blobby accretion" where over-dense blobs surpass the shock and thermalize in the atmosphere \citep{1988Frank}. This scenario requires at least L$_{soft}$/L$_{hard}$ $\sim$ 10 in  X-ray band and high H column densities. In our study, we do not  recover the soft component in the high state, and in both states column density of H is low. Thus, blobby accretion is unlikely.
In lower accretion rate states,  (${\dot{m}}\le1\ g~s^{-1} cm^{-2}$), it has been suggested that collisions in the shock heated gas may be inefficient and the Coulomb collisions of charged particles in the WD photosphere may be the way kinetic energy of the accreting gas is released \citep{2004Ramsay,1993Woelk}. Surveys of polars using \xmm\  and \swi\  show that  the soft X-ray components are absent in low states with detected hard X-ray plasma emission from the stand-off shock with temperatures $\le$ 5 keV at luminosities $\le 1\times10^{30}$\lumcgs \citep[][and references therein]{2004Ramsay,deMartinoetal2020}. The luminosities in the low state are $\sim$ a factor of 100 less than the high state \citep{2004RamsayA}. 
In general, polars change accretion state to a low state with a 3-4 magnitude difference in their V magnitude (Note here that this change could be 1.5-4 magnitudes if intermediate states are concerned). 

\srgcv\ changed to a low state at the beginning of 2021 from the high state in 2020  with a decrease of 3 mag. in system brightness. \nicer\ spectral results indicate that the neutral  H column densities are low. For the hard X-ray component in the low state, we detect a somewhat higher X-ray temperature in a range 7.4-13.5 keV when $\alpha$ parameter is not fixed (i.e., 0.3-0.7) as opposed to $\le\ 5$ keV generally measured in the low state polars. However, we note that the 3.0-6.0 keV range detected with $\alpha$=1 (at lower significance), is more consistent with the range derived in earlier polar surveys which assumed $\alpha$=1 in spectral modeling. \erosita\ data in the 0.2-9.0 keV range yield a factor of only $\sim$ 10-15 contrast between the high and low state fluxes for the hot plasma component (unabsorbed)  which is confirmed by the \nicer\ data as well.  We also found that the \nicer\ pulse profiles indicate energy dependence where the pulsed fraction below 1.4 keV is 2-3 times less than above 1.4 keV.  In the X-ray regime of polar-type CVs, occultations of the hotspot or the accretion column by the WD is an expected feature in the light curves producing modulations/variations \citep{2014Hellier,Mukai2017}. Photoelectric intrinsic absorption produces pulsed fraction that decreases with increasing energy in MCVs, mostly IPs, produced in the pre-shock accretion curtain \citep{1988Osborne,2014Hellier,Mukai2017}. The type of energy dependence we derive can not be properly explained in these context. We also do not find prominent eclipse effects in the optical or X-rays. \citet{1989Norton}, (cf. Fig. 4), calculate modulation depths in the X-rays as a result of a simple cold absorber or a patchy partial covering absorber where they find that the partial covering absorbers dilute the X-ray modulations and diminish/suppress the low energy (below 1 keV) spin pulsed-fractions. This is in accordance with our finding and hints at complex absorption scheme in our source, however, our
\nicer\ spectrum is not statistically adequate to perform modeling involving complex absorption. A decreased pulsed modulation in the lower energy band below $\sim$ 1-1.5 keV has been observed in cases like : V496 UMa \citep{2022Ok}, V1432 Aql \citep{2003Singh}, 1 RXS J2133.7$+$5107 \citep{2009Anzolin}, IGR J19552$+$0044 \citep{2013Bernardini}, and YY Dra \citep{2002Yazgan}. 

We detect the first soft X-ray component of a polar system in the low state with a blackbody temperature of 15-19 eV using \nicer\ data. This blackbody temperature is at the lower end of the blackbody temperature range of polars, 20-60 eV, determined from surveys \citep{deMartinoetal2020}.  \erosita\ data in the high state is consistent with the presence of the detected soft X-ray component by \nicer.  We calculated a stringent unabsorbed soft X-ray flux range performing fits with the SCORPEON variable background model (utilizing NICERDAS v.10a) using correct values from literature and fitting of the X-ray sky  background with its correct components to the \nicer\ spectrum (see also, Sec.~\ref{sec:ana-xray1}). We arrived at a best fitting temperature of  16.5$^{+1.3}_{-1.3}$ eV with a 2$\sigma$ error range of unabsorbed blackbody flux of  (2.8-20.3)$\times$10$^{-11}$ \fluxcgs\ in the 0.1-10.0 keV range (calculated using the contour plots of temperature and normalization around best-fit parameters). This energy range of soft X-ray blackbody flux underestimates the bolometric luminosity/flux by a factor of 18 times, only. 

The high state shows a dominant hot shocked-plasma component, kT$_{max} >$ 21 keV (lower limit at 95\% confidence level) with  a low N$_{\rm H}$ value of $0.6_{-0.3}^{+0.6}\times$10$^{21}$ cm$^{-2}$ and this value does not change in the state transition. The new MCV may be related to a recently proposed class of hard X-ray polars characterised with high X-ray temperatures and lack of the soft component in normal or high states \citep[e.g.,][]{2014Bernardini,2017Bernardini}. These systems are detected by \swi\ BAT and some with \integral. Presently, 13 out of 130 known polar systems are suggested as hard X-ray polars \citep{deMartinoetal2020,2019Bernardini}  and none has been found to show a soft X-ray component in any state as opposed to the one detected in this study using \nicer. 


\section{Summary and Conclusions}

We have optically identified a new polar-type magnetic cataclysmic variable discovered by SRG observatory during its all sky survey. The source was discovered in 2020 during a prominent flare, near its peak, and we observed it transmitting to the low state in 2021.  Optical and X-ray data were collected  both in high and low states of the source. 

We find the spin period of the WD (for standard polars is also the orbital period) to be 0.059710(1) d (85.982 min). We do not find any other periodicities in the optical or X-ray data. This period is one of the shortest (below the period gap) for polar-type systems. The orbital/spin pulse profile of the source is single peaked (mostly sinusoidal) in the high state, determined from the optical (TUG) data ($\sim$0.3 mag semi-amplitude of variations). In June 2021 (within the low state), the \nicer\ X-ray light curves shown in Fig.~\ref{Balman:nicerlc}, portray a double-peaked profile indicating a two-pole accretor. One of the maximum peaks is at phase $\sim$ 0.5, where the single peak in the optical exists during the high state (the accumulated phase error is $\sim$ 0.1 for \nicer\ data). 
However, optical photometry in the low state of the source in 2021  with Zeiss-1000 (October) and TUG (December) showed single-peaked pulse profile, similar to the one found in the optical band in the high state.  
We note that we do not have simultaneous optical and X-ray coverage. 
Change of accretion geometry and pole-switching from one to two-pole accretion within or across states is detected in several polar MCV systems \citep[e.g.,][]{2002Schwarz,2020Beuermann,2021Beuermann,2022Ok}.

We studied the spectrum of the new polar MCV during the high state in the optical (using TUG-RTT150 and SAO RAS-SCORPIO-1). Our low resolution spectroscopy revealed prominent Balmer lines with absorbed fluxes in a range (2.2-3.8)$\times$ 10$^{-14}$\ erg~s$^{-1}$~cm$^{-2}$~{\AA}$^{-1}$ and a prominent He{\sc ii}\ line with an absorbed flux of 2.8$\times$ 10$^{-14}$\ erg~s$^{-1}$~cm$^{-2}$~{\AA}$^{-1}$. The continuum does not reveal cyclotron humps in the high state. We have not been able to obtain long spectroscopic observations to find a well-constrained spectroscopic period, but it agrees with the photometric period. However, the asynchronism of the WD needs to be investigated. The equivalent widths of all the Balmer and He lines show variation over the photometric period of the WD and the detected EWs are consistent with recent results on polars \citep[cf. Table 5 in][]{2021Beuermann}. 

Our Study of Doppler tomography using the Balmer and He lines confirmed the polar MCV nature of the system. For polars ballistic section of the accretion flow is bounded by the region around the Lagrangian point L1 and the boundary defined by Alfven radius.  Increasing  accretion rate and decreasing magnetic field induction lengthen the ballistic trajectory \citep{2021Bisikalo}. Hence, our detection of the ballistic stream in the high state was to be expected for a polar system. This is also predicted to affect the morphology of the polar caps.

\srgcv\ was discovered in the X-rays by \artxc\ in the course of the SRG all-sky survey and followed-up with \nicer\ in the low state. \erosita\ data were contemporaneous with \artxc\ and provided accurate position and the spectral data of the X-ray source in the high state. One of the important findings of our study is the detection of the soft X-ray component in the \nicer\ data. This is the first detection of a soft component in the low state of a polar. Formal blackbody fits to the \nicer\ data give a temperature of 15-18 eV  with an unabsorbed flux of  (2.8-20.3)$\times$10$^{-11}$ \fluxcgs\ in the 0.1-10.0 keV band.
 
In the low state, we find that the plasma component remains hot around 7.4-13.5 keV (consistent with 3C50 and SCORPEON backrgound analysis) whereas in the high state the shocked plasma temperatures (in the PSR) are  kT$_{max} >$21 keV (lower limit at 95\% confidence level). The  low state and high state (hot plasma) unabsorbed fluxes are $\sim 4.0\times$ 10$^{-13}$\ erg~s$^{-1}$~cm$^{-2}$ and $\sim 6.0\times$ 10$^{-12}$\ erg~s$^{-1}$~cm$^{-2}$, respectively. This translates to about a factor of 10-15 decrease in unabsorbed flux for the hard X-ray (shock) emission spectrum in the 0.2-10.0 keV range from high into the low state which is about a factor of 5-10 times smaller than expected.  We find  a range of  low \nh, 0.5$\times$10$^{21}$--0.95$\times$10$^{21}$ cm$^{-2}$, which does not seem to change between high and low states. This value is about 5-10 times smaller than the Galactic absorption towards the source. It may suggest that the source is located in front of the main absorbing material in this direction and the source is closer by. Moreover, the intrinsic absorption may also be low, likely, partially covering at perhaps, relatively higher ionization level. Such complex absorption conditions can mimic low level of cold absorption in the X-ray spectra \citep[see also][]{2021Islam}.

\srgcv\ appears to have unique interesting characteristics. It may be related to a recently proposed group of hard X-ray polars (yet, these do not show high/low states changes). It may be an addition to a small number of low \nh\ polar systems  \citep[e.g.,][]{2019Bernardini,2018Webb,2008Vogel}. It also is the only polar MCV with a soft X-ray component detected in the low state. It hints at complex, patchy, and ionized intrinsic absorbing media and geometry changes across state transition. 

\section*{Acknowledgements}
We acknowledge T\"UBITAK National Observatory (TUG), IKI, KFU, and AST  for a partial support in using RTT150 (Russian-Turkish 1.5-m telescope in Antalya) with SRG \artxc\ project numbers 1805 and 1824. We thank the \nicer\ Observatory for performing the X-ray and T\"UBITAK National Observatory (TUG) and SAO  RAS (Special Astrophysical Observatory of the Russian Academy of Sciences) for the optical observations. We thank the \nicer\ calibration team, particularly Ron Remillard, for checking the total background; the low and high energy noise characteristics of our observation and confirming the soft excess in the data. We acknowledge 
ATLAS, ZTF,  and $Gaia$ for the availability of their data and light curves. HHE thanks TUG for partial support with project number 16ARTT150-949. The analysis of \erosita\ X-Ray  data  has been supported  by RSF grant N 23-12-00292. This study was partially funded by RFBR, project number 19-32-60048. EI and NS was partially supported by the subsidy FZSM-2023-0015 allocated to the Kazan Federal University. 

This work is based on data from Mikhail Pavlinsky \artxc\ and \erosita\ X-ray telescopes aboard the SRG orbital observatory. The SRG observatory was built by Roskosmos in the interests of the Russian Academy of Sciences represented by its Space Research Institute (IKI) in the framework of the Russian Federal Space Program, with the participation of the Deutsches Zentrum für Luft- und Raumfahrt (DLR). The \erosita\ X-ray telescope was built by a consortium of German Institutes led by MPE, and supported by DLR. The SRG spacecraft was designed, built, launched and is operated by the Lavochkin Association and its subcontractors. 
The \erosita\ data used in this work were processed using the eSASS software system developed by the German \erosita\ Consortium and proprietary data reduction and analysis software developed by the Russian \erosita\ Consortium. The \artxc\ and \erosita\ teams thank the Russian Space Agency, Russian Academy of Sciences and State Corporation Rosatom for the support of the SRG project and \artxc\ telescope and the Lavochkin Association (NPOL) with its partners for the creation and operation of the SRG spacecraft.





\bibliographystyle{aa}






\label{lastpage}
\end{document}